\documentclass[prb,reprint,amsmath,amssymb,aps,twocolumn,superscriptaddress]{revtex4-2}

\usepackage{hyperref}
\usepackage{rotating}
\usepackage{graphicx}
\usepackage{latexsym}
\usepackage{amssymb}
\usepackage{amsfonts}
\usepackage{soul}
\usepackage{color}
\usepackage{amsmath}
\usepackage{color}
\usepackage{lipsum}
\usepackage{natbib}
\usepackage{lineno}
\usepackage{graphicx}
\usepackage{xcolor}
\usepackage{ulem}

\begin{document}
\title{Comparing energy dissipation mechanisms within the vortex dynamics of gap and gapless nano-sized superconductors}

\author{E. C. S. Duarte}
\affiliation{Superconductivity and Advanced Materials Group, Departamento de F\'{\i}sica e Qu\'{\i}mica, Faculdade de Engenharia, \textit{Univ Estadual Paulista--UNESP} -Caixa Postal 31, 15385-000, Ilha Solteira, SP, Brazil.}

\author{E. Sardella}
\affiliation{Superconductivity and Advanced Materials Group, Faculdade de Ci\^{e}ncias,  \textit{Univ Estadual Paulista--UNESP}, Departamento de F\'{i}sica - Caixa Postal 473, 17033-360, Bauru, SP, Brazil}

\author{T. T. Saraiva}
\affiliation{HSE University, Moscow 101000, Russia}

\author{A. S. Vasenko}
\affiliation{HSE University, Moscow 101000, Russia}
\affiliation{I.E. Tamm Department of Theoretical Physics, P.N. Lebedev Physical Institute, Russian Academy of Sciences, 119991 Moscow, Russia}

\author{R. Zadorosny}
\email{rafazad@gmail.com}
\affiliation{Superconductivity and Advanced Materials Group, Departamento de F\'{\i}sica e Qu\'{\i}mica, Faculdade de Engenharia, \textit{Univ Estadual Paulista--UNESP} -Caixa Postal 31, 15385-000, Ilha Solteira, SP, Brazil.}

\begin{abstract}
The presence of magnetic fields and/or transport currents 
can cause penetration of vortices in superconductors. 
Their motion leads to dissipation and resistive state arises, 
which in turn strongly affects the performance of 
superconducting devices such as single-photon and 
single-electron detectors. Therefore, an understanding of the 
dissipation mechanisms in mesoscopic superconductors is 
not only of fundamental value but also very important for 
further technological advances. In the present work, we 
analyzed the contributions and interplay of the 
dissipative mechanisms due to the locally induced electric 
field and an intrinsic relaxation of the superconducting 
order parameter, $\Psi$, in mesoscopic samples by using the 
time-dependent Ginzburg-Landau theory. Although often 
neglected, we show that the dissipated energy due to 
relaxation of $\Psi$ must be taken into account for an 
adequate description of the total dissipated energy. The 
local increase of the temperature due to vortex motion 
and its diffusion in the sample were also analyzed, where 
the joint effect of thermal relaxation and 
vortex dynamics plays an important role for the dissipative 
properties presented by the superconducting systems.
\end{abstract}

\maketitle


\section{Introduction}
Mesoscopic superconductors present a variety of novel phenomena due to the confinement effects  experienced by supercurrents and vortices. Such downsized materials have found  applications in nanodevices like Superconducting Single-Photon Detectors (SSPDs) \cite{gol2001picosecond,kerman2007constriction,dorenbos2008low,Hadfield2009,berdiyorov2012spatially,zotova2012photon,Natarajan2012,gaudio2014inhomogeneous,renema2015effect} and Superconducting Single-Electron Detectors (SSEDs) \cite{rosticher2010high}. In both types of devices, a hot spot is created in the region where the particle (photon or electron) impacts the material and subsequently the heat diffuses through the material causing a local suppression of superconductivity. As a result, a spike of the measured voltage occurs caused by the burst of normal currents and the particle can thereby be detected~\cite{Semenov2002}. It is therefore evident that the understanding of both the electrical and thermal properties of mesoscopic superconductors is very important in such and similar devices \cite{Kozerov2000,yang2007modeling}.

When an external magnetic field is applied in a type II superconductor above the so-called ``penetration field'', $H_{c1}$, magnetic flux in the form of Abrikosov vortices pierce the sample and the material transits into a mixed state~\cite{Abrikosov1957a,Brandt1995}. In the presence of electric current, the motion of those vortices dissipates energy, which results in a resistive state~\cite{TinkhamBook}. Several models have considered only the electric field induced by the moving vortices as a dissipation mechanism, which generates heat by the Joule effect \cite{berdiyorov2012spatially,berdiyorov2012magnetoresistance,berdiyorov2012large,hernandez2008dissipation}. In this respect, Bardeen and Stephen described the connection between the vortex velocity and the induced electric field by considering a viscous motion proportional to the square velocity of the vortex \cite{bardeen1965theory}. However, such a mechanism predicted much smaller damping than experimentally observed. Later on, Tinkham \cite{tinkham1964viscous} proposed the existence of another dissipation mechanism, which is related to an intrinsic relaxation of the order parameter, $\Psi$.
During the 90s and 2000s, the microscopic nature of the dissipative mechanisms and their influence on the viscosity and resistivity have been extensively studied experimentally up to recent years~\cite{Gerhenzon1984,Gershenzon1990,Marsili2016,Sheikhzada2020} with some experimental achievements~\cite{Tanaka1996,Bitauld2010}.

Despite that, there is still the need for more refined studies about how macroscopic parameters (within the time-dependent Ginzburg-Landau theory) affect the dissipation mechanism and, consequently, how they influence the thermal diffusion process. Also, due to the mesoscopic scale of these detectors, sample geometry plays a fundamental role yet not deeply addressed. By analyzing these issues, we present in this work a detailed mapping of the resistive state for nanoscale sizes square superconductors, with a focus on the thermal diffusion process during the Abrikosov's vortex motion using the time-dependent Ginzburg-Landau formalism.

The next sections of this article are organized as follows: in Sec.~\ref{sec2}, we outline our theoretical methods applied in order to obtain our results. In Sec.~\ref{sec3}, we present our studies of the influence of the diffusion parameter, ${u}$, and the confinement effects on the resistive state for gap and gapless superconductors. In Sec.~\ref{sec4}, we focus on the behavior of the dissipative mechanisms, thermal diffusion and non-equilibrium aspects of the magnetization and voltage for both gap and gapless superconductors. In the last part, Sec.~\ref{disconc} we present the discussion of our results and the conclusions.

\section{Theoretical formalism}\label{sec2}

The time-dependent approach for the Ginzburg-Landau (TDGL) equations, which were firstly proposed by Schmid \cite{schmid1966time}, provides a temporal evolution of the order parameter $\Psi$ and the vector potential $\textbf{A}$ for a superconducting material submitted to an external applied magnetic field and/or a transport current for gapless superconductors. Later on, Kramer and Watts-Tobin presented a generalized version for the TDGL equations, which took into account the superconducting gap \cite{kramer1978theory}. Both phenomenological non-equilibrium approaches are appropriate to describe most phenomena that occurs in the resistive state. The free energy theorem concerning the dissipative mechanism was first derived by Schmid~\cite{schmid1966time}. In Reference~\cite{duarte2017dynamics}, we generalized this theorem for gap superconductors. In some special circumstances, such theorem is in agreement with experimental observations, even for low temperatures \cite{petkovic2016deterministic}. Strictly speaking, both time dependent versions of the GL equations (either gapless or gap superconductors) are valid in the dirty limit and very close to ${T}_{c}$. Despite this limitation of Kramer-Watts-Tobin extension of the TDGL theory, some agreements with experiments can be achieved at temperatures as lower as ${T} = 0.5{T}_{c}$ \cite{petkovic2016deterministic,schmid1966time}. Those equations have also been applied in studies with induced voltage \cite{vodolazov2005masking}, magnetoresistance \cite{berdiyorov2012magnetoresistance,berdiyorov2012large}, kinematic vortices \cite{silhanek2010formation}, flux-flow regime \cite{jelic2015stroboscopic,jelic2016velocimetry} and for samples under AC external magnetic fields \cite{hernandez2008dissipation}. The generalized time dependent Ginzburg-Landau (GTDGL) equations in dimensionless units 
are given by:
\begin{align}
&\frac{{u}}{\sqrt{{1+\gamma^2|\Psi|^2}}}\Bigg(\frac{\partial}{\partial{t}} + {i}\varphi
+ \frac{1}{2}\gamma^2\frac{\partial|\Psi|^2}{\partial{t}}\Bigg)\Psi=\nonumber\\
&\qquad\qquad (\nabla - i\textbf{A})^2\Psi + (1 - {T} - |\Psi|^2)\Psi\label{eq:1}
\end{align}
and
\begin{equation}
\Bigg(\frac{\partial\textbf{A}}{{\partial{t}}} + \mbox{\boldmath $\nabla$}\varphi\Bigg) = {\textbf{J}}_{s} - \kappa^2\mbox{\boldmath $\nabla$}\times\mbox{\boldmath $\nabla$}\times{\textbf{A}},
\label{eq:2}
\end{equation}
where the superconducting current density is
\begin{equation}
{\textbf{J}}_{s} = {\rm Im}\left[\Psi^\ast(\mbox{\boldmath $\nabla$}-i{\textbf{A}})\Psi\right].
\label{eq:3}
\end{equation}

Here, the order parameter is expressed in units of its bulk value at zero external field, $|\Psi_0|^2=8\pi^2T_c^2/7\zeta(3)$, the distances are expressed in units of the coherence length at zero temperature $\xi(0)^2=\frac{7\zeta(3)}{48\pi^2T_c^2}\hbar^2v_F^2$, where $v_F$ is the Fermi velocity, and the magnetic field is expressed in units of the bulk upper critical field $H_{c2}(0)=\Phi_0/2\pi\xi^2$, where $\Phi_0=hc/2e$ is the quantum unit of flux. The temperature is expressed in units of ${T}_{c}$, time is in units of ${t}_{GL}(0) = \pi\hbar/8{k}_{B}{T}_{c}{u}$, the Ginzburg-Landau characteristic time; $\varphi$ is the scalar potential and it is expressed in units of $V_0=\hbar/2et_{GL}(0)$. The GL parameter is $\kappa=\lambda(0)/\xi(0)$, where $\lambda(0)$ is the London penetration depth at zero temperature. The parameter $\gamma = 2\Delta(0)\tau_{e-ph}/\hbar$ is related to the gap of the superconductor at zero temperature $\Delta(0)\approx1.8k_BT_c$, where $\tau_{e-ph}$ is the electron-phonon inelastic collision time and $k_B$ is the Boltzmann constant. The parameter ${u}$ is given by the microscopic derivation of the Ginzburg-Landau equations. For a gapless superconductor, $\gamma = 0$, Gor'kov and Eliashberg obtained ${u} = 12$ \cite{gor1975vortex}. In this case, Kramer and Baratoff \cite{kramer1977lossless} define ${u} = \frac{\tau_{|\Psi|}(0)}{t_{GL}(0)}$, with $\tau_{|\Psi|}(0)$ being the relaxation time of $|\Psi|$. On the other hand, for gap superconductors, ${u} = 5.79$ as demonstrated by Kramer and Watts-Tobin \cite{kramer1978theory}. Several numerical algorithms have been developed to solve the TDGL equations numerically (see for instance References~\cite{gropp1996numerical,milovsevic2010ginzburg}). We have used the link-variable method  which ensures the gauge invariance when such equations are discretised on a numerical grid, as shown in~\cite{gropp1996numerical}. For all times and positions we have chosen $\varphi = 0$, since neither charges nor external currents are considered in this work. To obtain the total dissipated power density, we generalized the Helmholtz free energy theorem proposed in the Reference~\cite{schmid1966time} for a superconductor in an external magnetic field. Such equation, in dimensionless form, is given by:
\begin{eqnarray}
{W}_{Total} &=& 2\bigg(\bigg|\frac{\partial\textbf{A}}{\partial{t}}\bigg|\bigg)^{2} \\ \nonumber
& & + \frac{{2u}}{\sqrt{{1+\gamma^2|\Psi|^2}}}\Bigg[\Bigg|\frac{\partial\Psi}{\partial{t}}\Bigg|^2 
+ \frac{\gamma^{2}}{4}\Bigg(\frac{\partial|\Psi|^2}{\partial{t}}\Bigg)^2\Bigg].
\end{eqnarray}

We distinguish the ${W}_{Total}$ in the following 
dissipative mechanism:

\begin{align}
{W}_{\textbf{A}} =& 2\bigg(\bigg|\frac{\partial\textbf{A}}{\partial{t}}\bigg|\bigg)^{2}, \label{eq.disA}\\
{W}_{\Psi} =&\frac{{2u}}{\sqrt{{1+\gamma^2|\Psi|^2}}}\Bigg[\Bigg|\frac{\partial\Psi}{\partial{t}}\Bigg|^2
\Bigg],\label{eq.dispsi} \\
{W}_{\gamma} =& \frac{{2u}}{\sqrt{{1+\gamma^2|\Psi|^2}}}\Bigg[\frac{\gamma^{2}}{4}
\Bigg(\frac{\partial|\Psi|^2}{\partial{t}}\Bigg)^2\Bigg].\label{eq.disgam}
\end{align}
The extension of the Schmid theorem was first extended to the GTDGL equation in \cite{duarte2017dynamics}.

The first term is the dissipation due to the induced electric field, ${W}_{\textbf{A}}$, the second one is related to the relaxation of the order parameter, ${W}_{\Psi}$, and the last term is due to the relaxation of the density of superelectrons, ${W}_{\gamma}$. The dissipated power density is given in units of $(a_0T_{c})^2/{b}{t}_{GL}(0)$, where ${a}_{0}$ and ${b}$ are the phenomenological GL parameters \cite{sardella2006temperature}. As the dissipated energy diffuses through the sample, we coupled the thermal diffusion equation to the GTDGL ones. By using the approach of Ref.~\cite{vodolazov2005masking}, the dimensionless form of the thermal equation takes the form:
\begin{eqnarray}
{C}_{eff}^\prime\frac{\partial{T}}{\partial{t}} =  {K}_{eff}\mbox{\boldmath$\nabla$}^2{T} +
\frac{1}{2}{W}_{Total} - \eta({T} - {T}_{0}),
\label{eq:5}
\end{eqnarray}
where $\eta$ is the heat-transfer coefficient to the substrate, ${T}_{0}$ is the bath temperature, ${C}_{eff}^\prime = \pi^4/48{u}$ is the effective heat capacity, and ${K}_{eff} = \pi^4/48{u}^2$ is the effective thermal conductivity. As one can see, the parameter ${u}$ influences the behavior of the thermal proprieties as well.

To minimize the time spent by the simulations, several works used ${u} \leq 1$ without losing qualitative correspondence to experimental data \cite{berdiyorov2012magnetoresistance,vodolazov2000effect}. However, the right values of ${u}$ must be used in studies of non-equilibrium processes such as thermal dissipation and diffusion. To have a proper correspondence to experimental data, the GTDGL equations must follow the condition $ 1-T < \frac{u^{1/2}}{\gamma}$ \cite{tidecks1986continuous}. On the other hand, for $\gamma = 0$, we recover the TDGL equations, which are derived for the case of a large a concentration of paramagnetic impurities, i.e., ${u} = 12$ and ${T} \rightarrow {T}_{c}$ or (${H} \rightarrow {H}_{c2}$) \cite{gorkov1971viscous}. References~\cite{baranov2011current,ivlev1980dynamics,ivlev1985low} report such modifications to properly describe the gapless superconductors.

\section{Influence of the diffusion parameter $u$ over the dissipation resistive state} \label{sec3}

\begin{figure}[!t]
    \begin{center}
    \includegraphics[width=0.8 \linewidth]{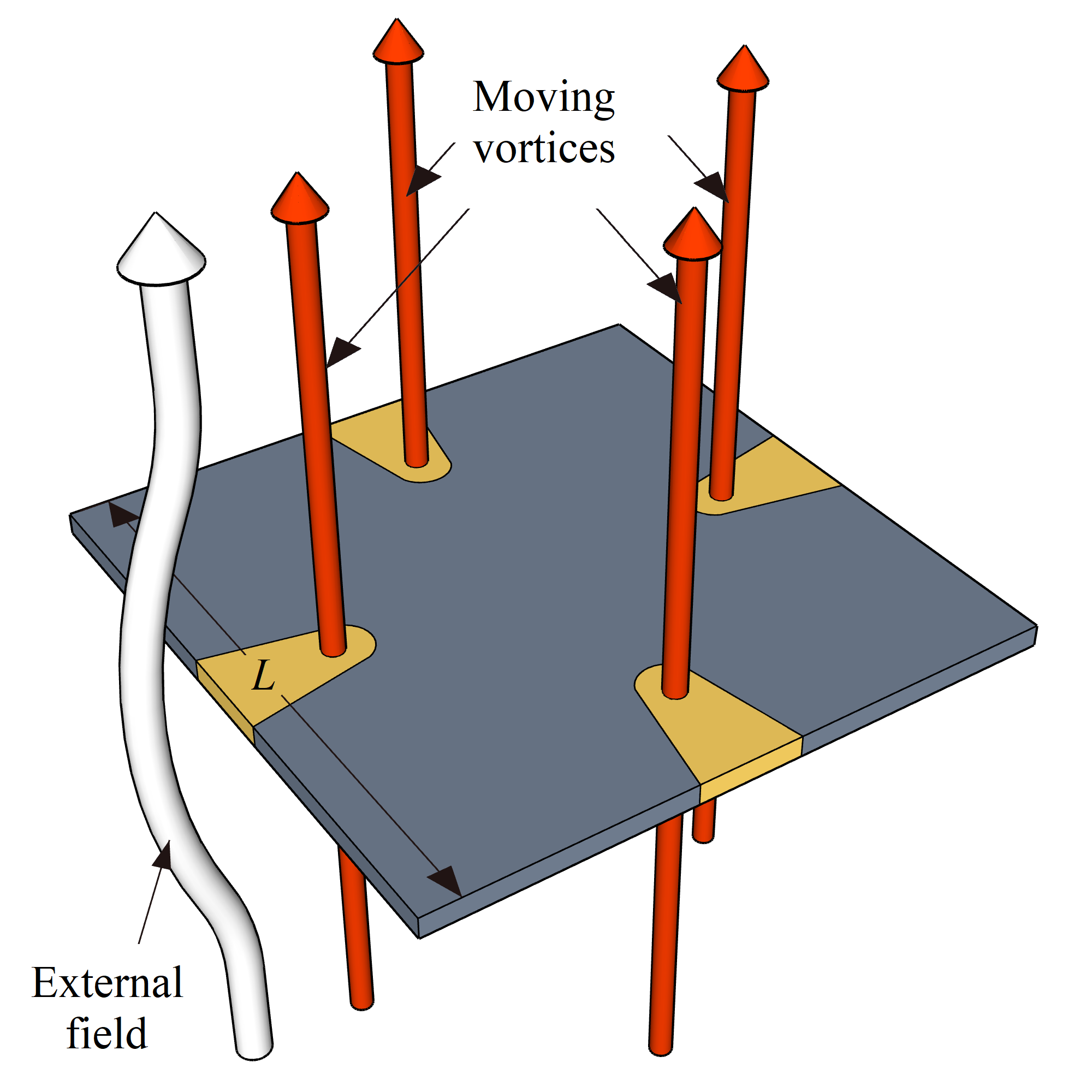}
    \caption{Sketch of the square superconducting thin film submitted to a perpendicular external magnetic field. Here we show the moment when four quantum of fluxes penetrate the sample dissipating heat and causing a local increase in temperature, represented by the yellow region.}
    \label{fig:sketch}
    \end{center}
\end{figure}

We carried out numerical simulations for square thin films with lateral sizes ranging from  ${L} = 10\xi(0)$ to $100\xi(0)$, such as sketched in Fig.~\ref{fig:sketch}.
We employed the superconductor/insulator boundary condition, i.e. the normal component of the superconducting current at the boundary is zero: $\textbf{n}\cdot(\mbox{\boldmath$\nabla$} - i\textbf{A})\Psi = 0$, and also we assumed that all the heat is transferred to the substrate, which means $\textbf{n}\cdot(\mbox{\boldmath $\nabla$}{T}) = 0$. In all cases, the mesh was set as ${\Delta x} = {\Delta y} = 0.1\xi(0)$, the GL parameter $\kappa = 5$ and the bath temperature ${T}_{0} = 0.93{T}_{c}$. The external magnetic field was varied in steps of $\Delta{H} = 10^{-3}{H}_{c2}(0)$ and the heat-transfer coefficient $\eta$ was set as $2\times 10^{-4}$. This is an intermediate value for heat removal as  discussed in Reference~\cite{vodolazov2005masking}. The density of power dissipation was analyzed during the first vortex penetration by setting the diffusion parameter ${u}$ equal to $1$, $5.79$ and $12$.

\begin{figure}[t]
    \begin{center}
    \includegraphics[width=\linewidth]{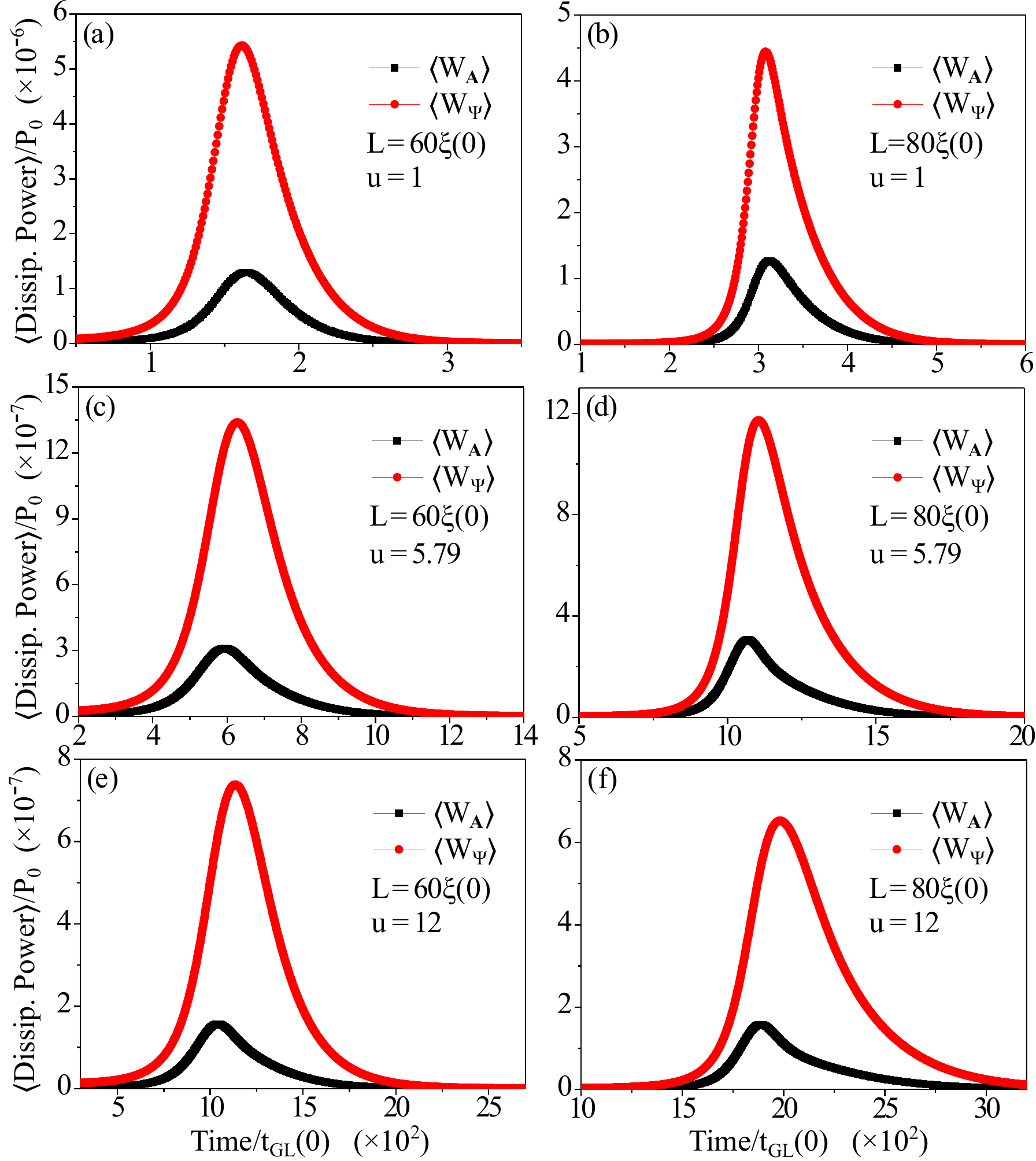}
    \caption{Average dissipated power density as a function of time during the first vortex penetration for samples with dimensions ${L} = 60\xi(0)$ (left column) and ${L} = 80\xi(0)$ (right column) for some values of ${u}$ and ${T}_{0} = 0.93{T}_{c}$.}
    \label{fig:Figure2}
    \end{center}
\end{figure}

The evolution of the spatial averages of the contributions to dissipation, $\langle{W}_{\textbf{A}}\rangle$ and $\langle{W}_{\Psi}\rangle$, is shown in Fig.~\ref{fig:Figure2} for samples with sides $L=60\xi(0)$ and $L=80\xi(0)$ for $u=\{1,\ 5.79,\ 12\}$.
As one can notice, increasing the parameter $u$ does not influence the qualitative behavior of the system, but can decrease the dissipation around one order of magnitude. Also, the total duration of the event is longer for the largest sample, $L=80\xi(0)$. Besides that, the maximum of the dissipation is reached faster for smaller sample. Such behavior occurs when the lateral size of the sample is near to the meso-macro crossover according to the criterion used in Ref.~\cite{zadorosny2012crossover}.

\begin{figure}[t]
	\centering\vspace*{0.3cm}
	\includegraphics[width=\linewidth]{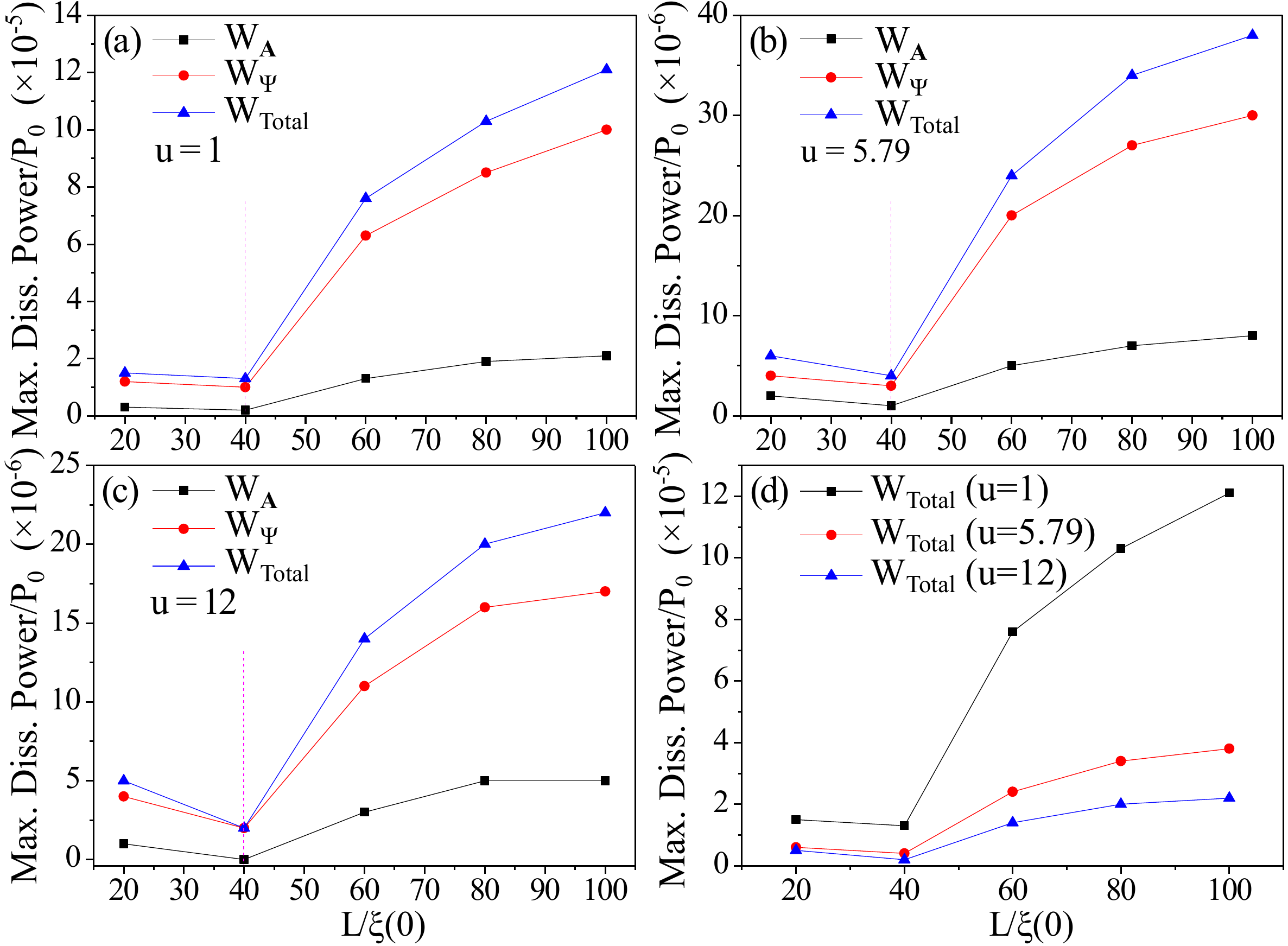}
	\caption{\label{fig:Figure3}Local maximum of the dissipated power density for 
	different values of ${u}$ as a function of lateral size ${L}$ at ${T}_{0} = 0.93{T}_{c}$. 
	The dashed line indicates a vorticity transition from 2 to 4 vortices.}
\end{figure}

In Fig.~\ref{fig:Figure3} it is shown the maximum values reached by ${W}_{\textbf{A}}$, ${W}_{\Psi}$ and ${W}_{total}={W}_{\textbf{A}}$ + ${W}_{\Psi}$ as functions of the lateral size of the samples, for the first vortex penetration.
We can observe that all dissipative mechanisms increase for $L\ge 40\xi(0)$, where the first penetration corresponds to four or more vortices. For such values of $L$, the vortices have more space to move freely before their repulsive interaction becomes more pronounced. Such a fact explains the increasing dissipation for larger values of $L$. On the other hand, confinement effects for ${L} = 20\xi(0)$ are responsible for the penetration of only two vortices. Then, the repulsive interaction is present since the beginning of their penetration (see Reference~\cite{zadorosny2012crossover}). Besides that, the shielding current produces a greater Lorentz force over the two penetrated vortices making their dissipation larger in comparison with ${L} =40\xi(0)$. From Fig.~\ref{fig:Figure3}(d) we can also notice a saturation tendency of the dissipative mechanisms curves when the transition from the mesoscopic to the macroscopic regime is reached, which occurs around ${L} = 80\xi(0)$ for ${T}_{0} = 0.93{T}_{c}$. It is remarkable that this value is very closed to the one predicted in Ref.~\cite{zadorosny2012crossover}. Furthermore, in Fig.~\ref{fig:Figure2}(d) it was obtained ${W}_{total}^{{u}=5.79} \simeq {W}_{total}^{{u}=12}$ with $20\xi(0) \le {L}\le 40\xi(0)$. As we pointed out in Sec.~\ref{sec2}, those cases were derived from similar approaches for both values of ${u}$, i.e., derived considering the dirty limit of a microscopic theory. This result corroborates the consistency of our models with the microscopic theory. In short, for small superconducting systems, where strong confinement effects drive the vortex dynamics, there is no physical difference in describing those systems by using either $u=12$ or $u=5.79$ when the dissipated power is taken into account.

\begin{figure}[t]
	\centering\vspace*{0.1cm}
	\includegraphics[width=\linewidth]{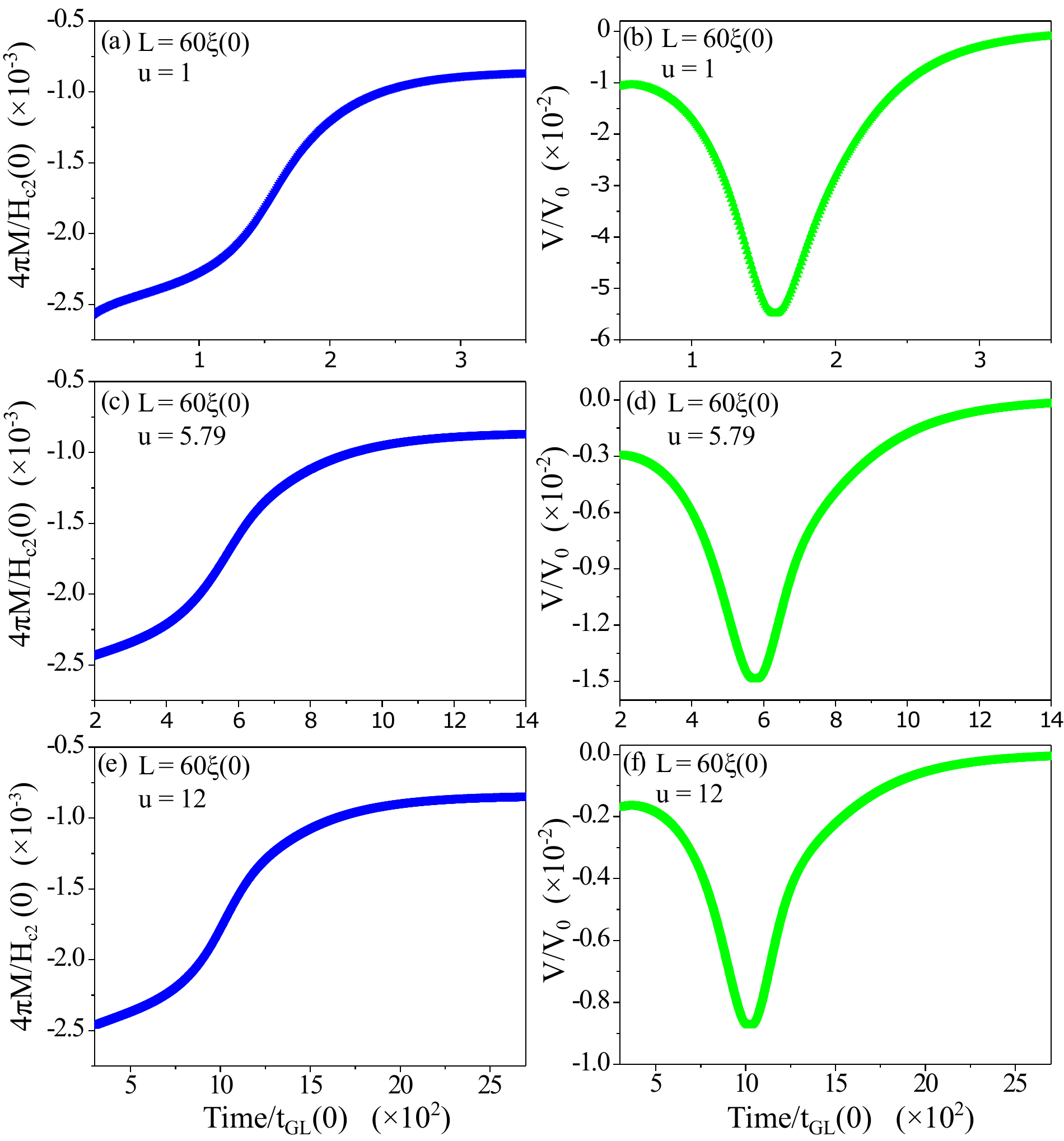}
	\caption{\label{fig:Figure4}Magnetization (left column) and voltage (right column) as functions of time during the vortex penetration for different values of ${u}$.}
\end{figure}

Concluding this section, we analyzed the influence of ${u}$ over the magnetization and the voltage pulse induced by the first flux penetration (see Fig.~\ref{fig:Figure4}).
One can notice that the characteristic time of the whole event grows almost linearly with $u$ when expressed in units of $t_{GL}$. We obtained characteristic times around $350t_{GL}(0),\ 1200t_{GL}(0)$ and $2500t_{GL}(0)$ for $u=1,\ 5.79,\ 12$, respectively. But, in general, one can express these time values in seconds as $t_{GL}\approx3\times10^{-12}/T_c u$. Using the parameters for WSi from Ref.~\cite{Zhang2018}, $T_c=3.5K$, this means that the characteristic times for this penetration of vortices happen within $\sim300$ps (with $u = 1$), $\sim178$ps (with $u = 5.79$) and $\sim179$ps (with $u = 12$). We believe that $u$ should be high for amorphous samples. The amplitude of the voltage pulse decreases with $u$ when expressed in units of $V_0$: $5.5\times10^{-2}V_0,\ 1.5\times10^{-2}V_0$ and $0.88\times10^{-2}V_0$ for $u=1,\ 5.79,\ 12$. Again, we can consider the case of WSi and get the modulus of these voltage peaks between $\sim7.7\times10^{-8}statV$ for $u=1$, $\sim1.2\times10^{-7}statV$ for $u=5.79$ and $\sim1.5\times10^{-5}statV$ for $u=12$.

\section{Resistive state of vortex-motion in a multiply-connected sample}\label{sec4}
\begin{figure}[t]
	\centering\vspace*{0.1cm}
	\includegraphics[width=0.9\columnwidth]{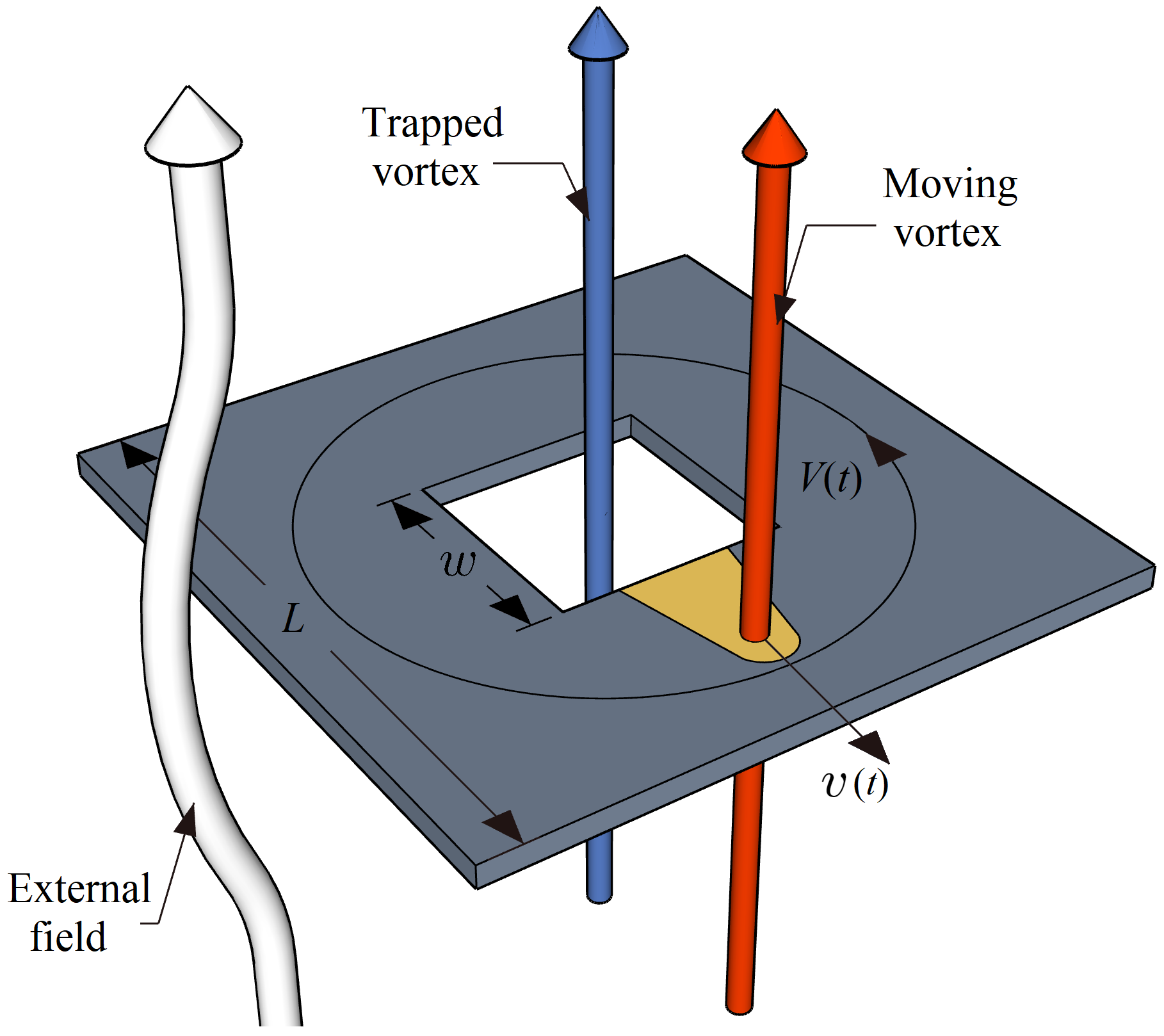}
	\caption{\label{sketch_b}Sketch of the thin square superconducting film with an square occlusion in the center with side $w$. After trapping two vortices in the sample, one can decrease the external field (represented by a white arrow) and one of the fluxes leave the sample with velocity $v$ and creating a ``hotspot''. The blue arrow represents the trapped vortex that remains in the sample and the red one represents the flux quantum that leaves the sample. During this process, a non-zero voltage, $V(t)$, around the sample can be measured.}
\end{figure}

In order to study the dissipative mechanisms during the motion of a single-vortex in detail, we simulated square superconducting samples with size ${L}$ and a central square hole (or antidot, AD), with side ${w}$. The applied magnetic field was increased in steps of $\Delta H = 10^{-3}H_{c2}(0)$ up to the penetration of two vortices and their consequent trapping in the AD. After that, the applied magnetic field was decreased until the depinning of one vortex, as sketched in Fig.~\ref{sketch_b}.

\subsection{Gapless superconductors}\label{subsec.gapless}
\begin{figure}[t]
 	\centering\vspace*{0.3cm}
 	\includegraphics[width=\linewidth]{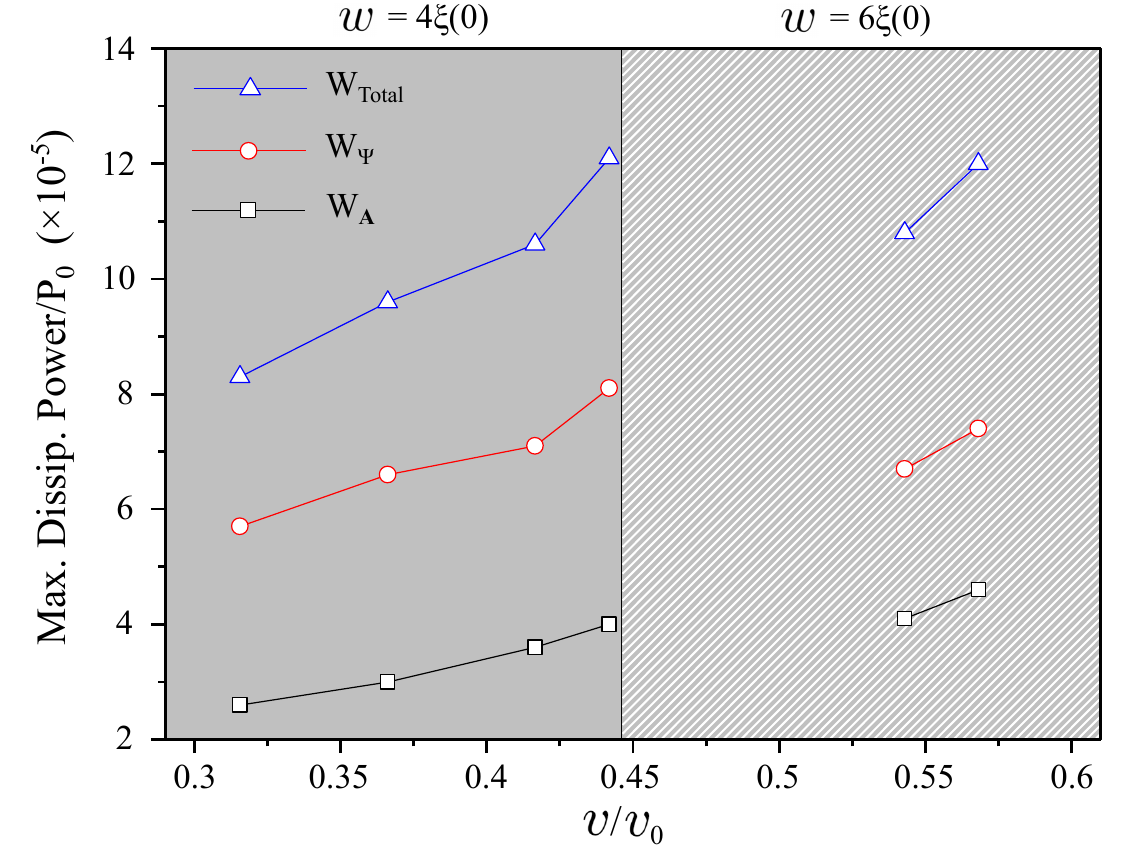}
 	\caption{\label{fig:Figure6}Relation between the local maximum of the dissipated power and the vortex velocity, $v$, for a gapless superconductor (i.e. $\gamma=0$) at $T=0.93T_c$. The gray region contains the results for the sample with ${w} = 4\xi(0)$ while the dashed stands for ${w} = 6\xi(0)$.}
\end{figure}

\begin{figure}[t]
  \centering\vspace*{0.3cm}
  \includegraphics[width=\linewidth]{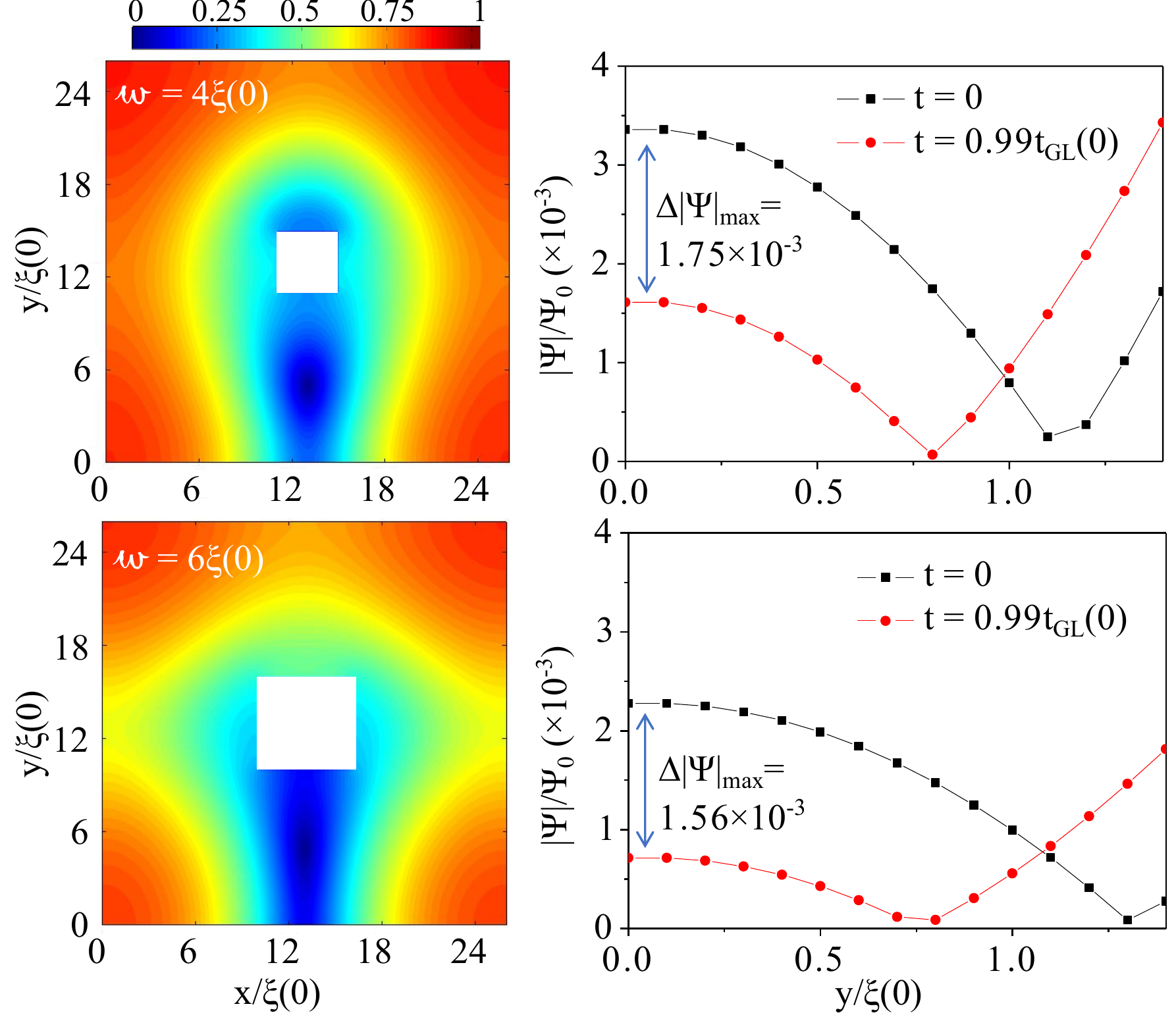}
  \caption{\label{fig:Figure7}On the left, the intensity of $|\Psi|$ illustrates the changing on the vortex shape for two samples of size ${L} = 26\xi(0)$, for ${w} = 4\xi(0)$ (upper left panel) and ${w} = 6\xi(0)$ (lower left panel). In both panels the core of the vortex is localized at ${y} = 5.0\xi(0)$. The right panels show the variations of $|\Psi|$ along the vortex core just before the vortex leaves the sample in a time interval defined from ${t} = 0$ to ${t} = 0.99{t}_{GL}(0)$ along the vertical line ${x} = 13\xi(0)$, i.e. the middle of the sample.}
\end{figure}

The process of a quantum flux leaving the sample has many aspects to be investigated. For example, the current induced by the spike in the voltage has to flow partly through the center of the vortex, which is a normal state region and this causes dissipation. In Fig.~\ref{fig:Figure6}, it is shown the local maxima of ${W}_{\mathbf{A}}$, ${W}_{\Psi}$, and ${W}_{total}$ as functions of the vortex velocity when moving out of the sample, $v$.
In order to investigate how the relation between the dissipative mechanisms and the vortex velocity depends on the geometry, we fixed the dimension of the square ${L=26\xi(0)}$ and considered two different values of the AD size, ${w}$. On the left-hand side of Fig.~\ref{fig:Figure6} one finds the data for ${w} = 4\xi(0)$ and on the right-hand side for ${w} = 6\xi(0)$. In general, one can see that the dissipated power increase monotonically as functions $v$, as expected. But the system with ${w} = 6\xi(0)$ shows smaller values for both ${W}_{\Psi}$ and ${W}_{\bold A}$ which is related to the fact that the vortex is concentrated in a smaller region for ${w} = 4\xi(0)$. To understand such behavior, we plotted the intensity of $|\Psi|$ for those values of ${w}$ with ${L} = 26\xi(0)$, as illustrated in the panels of Fig.~\ref{fig:Figure7}. They show that the vortex core becomes more stretched in the $y$ direction as the velocity of the vortex increases, by forming a phase-slip-like region, similar to a \textit{vortex street}~\cite{berdiyorov2009kinematic}. This change in the vortex core shape affects the time-variation of $|\Psi|$ (see the plots in Fig.~\ref{fig:Figure7}). It can be seen that the maximum value of $\Delta|\Psi|$ is lower for the sample with ${w} = 6\xi(0)$ and consequently ${W}_{\Psi}$ decreases.
Also, these panels show that the modulus of the order parameter increases sharply after the vortex core in the direction to the boundary while it is slowly increasing in the direction to the hole in the center of the sample. In other words, the second derivative of the modulus of order parameter changes sign after crossing the vortex core.

\begin{figure}[t]
	\includegraphics[width=\linewidth]{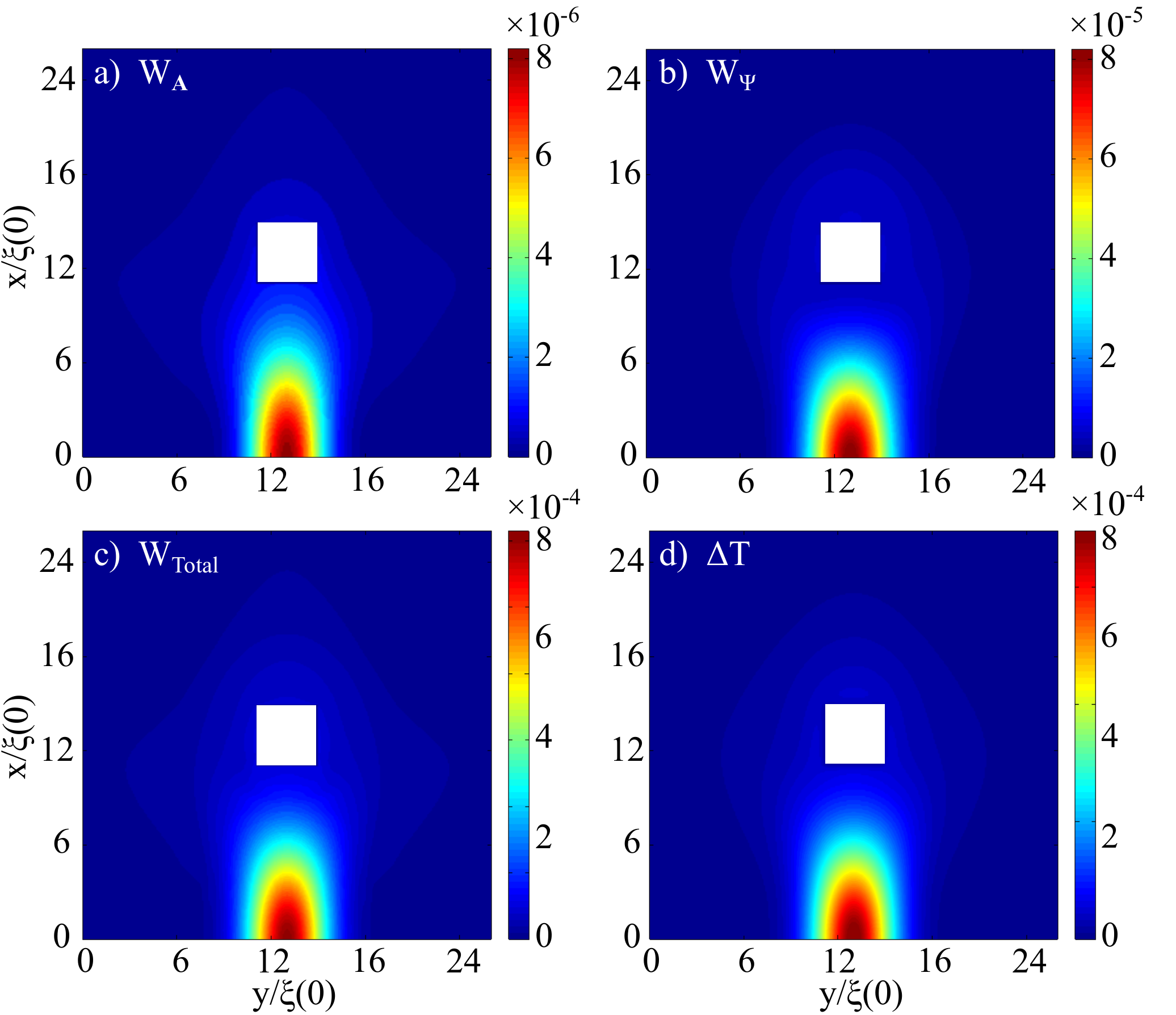}
	\caption{\label{fig:Figure8}Local intensity profile of the dissipated power functions and the local variation of temperatures for a gapless superconductor ($\gamma=0$) with parameters ${L} = 26\xi(0)$ with ${w} = 4\xi(0)$. The snapshot corresponds to the moment where one of the two trapped vortices is about to leave the sample. This is the moment of maximum dissipation and also the point of highest vortex velocity.}
\end{figure}

As expected, vortex motion is accompanied by dissipation of heat and a local increase of the temperature. In Fig.~\ref{fig:Figure8}, it is shown snapshots of the intensity of each dissipative mechanisms, and the variation of the local temperature, $\Delta{T} = {T} - {T}_{0}$, when the vortex is about to leave the sample. We chose the moment when $W_{Total}$ reaches its local maximum (also seen in Fig.~\ref{fig:Figure6} that it is the moment of fastest vortex velocity). The maximum dissipated power $W_\Psi$ shows around one order of magnitude larger than $W_\bold A$ being the main responsible for the local increase in temperature. One can see in Fig.~\ref{fig:Figure8} that $\Delta{T}$ follows directly the behavior of $W_{Total}$, i.e. it is larger when the vortex is about to leave at the lower edge of the sample and its velocity is maximum.
Furthermore, in these plots shown in Fig.~\ref{fig:Figure8}, the dissipation shows a particular behavior near the outer edges of the sample for each contribution. The term $W_{\bold A}$ is more pronounced near the edges of the sample~\cite{Brosens1999}, while $W_{\Psi}$ shows less influence by shading from the corners. This can be interpreted by the fact that the superconducting currents can better circulate near the corners while not so much near the edges and thus the stronger effect over $W_{\bold A}$. This influence from the corners affects the total dissipation but becomes slightly less visible in the temperature change $\Delta T$.

\subsection{Gap superconductor}\label{subsec.gap}
We also studied the dissipative mechanisms during the vortex motion in a gap superconductor, i.e. we considered $\gamma\neq0$ in Eq.~(\ref{eq:1}). For the sake of estimating its order of magnitude, one can use the BCS expression for the gap $\Delta(0)=1.76k_BT_c$ and substitute in the definition shown in the Sec.~\ref{sec2}, $\gamma=2\Delta(0)\tau_{e-ph}/\hbar$. Furthermore, based on the values obtained for WSi in Ref's.~\cite{Sidorova2018,Zhang2018,Zhang2019}, ranging from 75ps to 190ps, i.e. within the order $\tau_{e-ph}\sim10^{-10}$s with $T_c\sim10^{0}K$. This means that $\gamma\sim10^{1}$.

\begin{figure}[t]
	\centering\vspace*{0.3cm}
	\includegraphics[width=\linewidth]{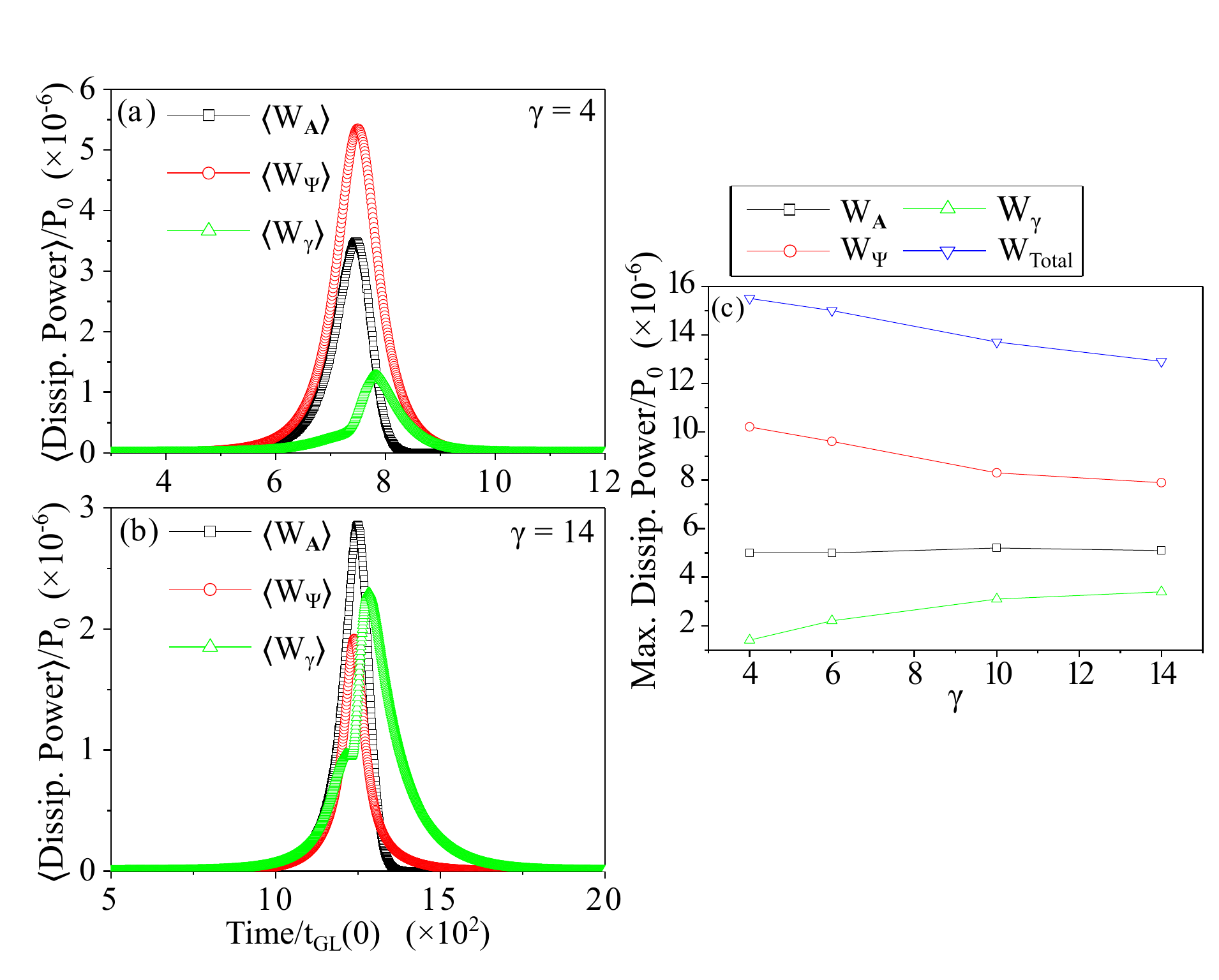}
	\caption{\label{fig:Figure9}Average dissipated power density as a function of  time for (a) $\gamma = 4$ and (b) $\gamma = 14$. In (c) it is shown the local maximum of the dissipated power density as a function of $\gamma$.}
\end{figure}

The setup of the simulation and shape of the samples were similar to the ones considered previously in Sec.~\ref{subsec.gapless}, i.e. thin square films with a square hole in the center, and the magnetic field was decreased until a vortex moves out of the sample. Along the whole set of parameters that we covered ($\gamma$, $L$, $w$), the extra dissipated power due to this term, $\langle W_\gamma\rangle$, was typically smaller or of the same order of magnitude than the other contributions, $\langle W_\mathbf{A}\rangle$ and $\langle W_\Psi\rangle$, as shown in Fig.~\ref{fig:Figure9} for ${L} = 26\xi(0)$, ${w} = 4\xi(0)$, and $u=5.79$.
One can see in panels (a) and (b) that $\langle{W}_{\gamma}\rangle$ increases with $\gamma$, whereas $\langle{W}_{\Psi}\rangle$ decreases. It is interesting to note that this is a direct consequence of their dependence on $\gamma$ in Eq's.~\ref{eq.disgam} and~\ref{eq.dispsi}. Also, as expected, $\langle{W}_{\mathbf{A}}\rangle$ is practically not affected, also expected from the definition in Eq.~\ref{eq.disA}. The total dissipation decreases with $\gamma$, indicating that the dissipation is stronger in gap superconductors for this particular setup. In addition, the local maximum power dissipation in panel~\ref{fig:Figure9}-(c) shows that this trend maintains or, in other words, ${W}_{\Psi}$ decreases while ${W}_{\gamma}$ increases smoothly and the maximum of ${W}_{total}$ decreases smoothly with increasing $\gamma$ as well.

\begin{figure}[h!]
	\centering\vspace*{0.3cm}
	\includegraphics[width=\linewidth]{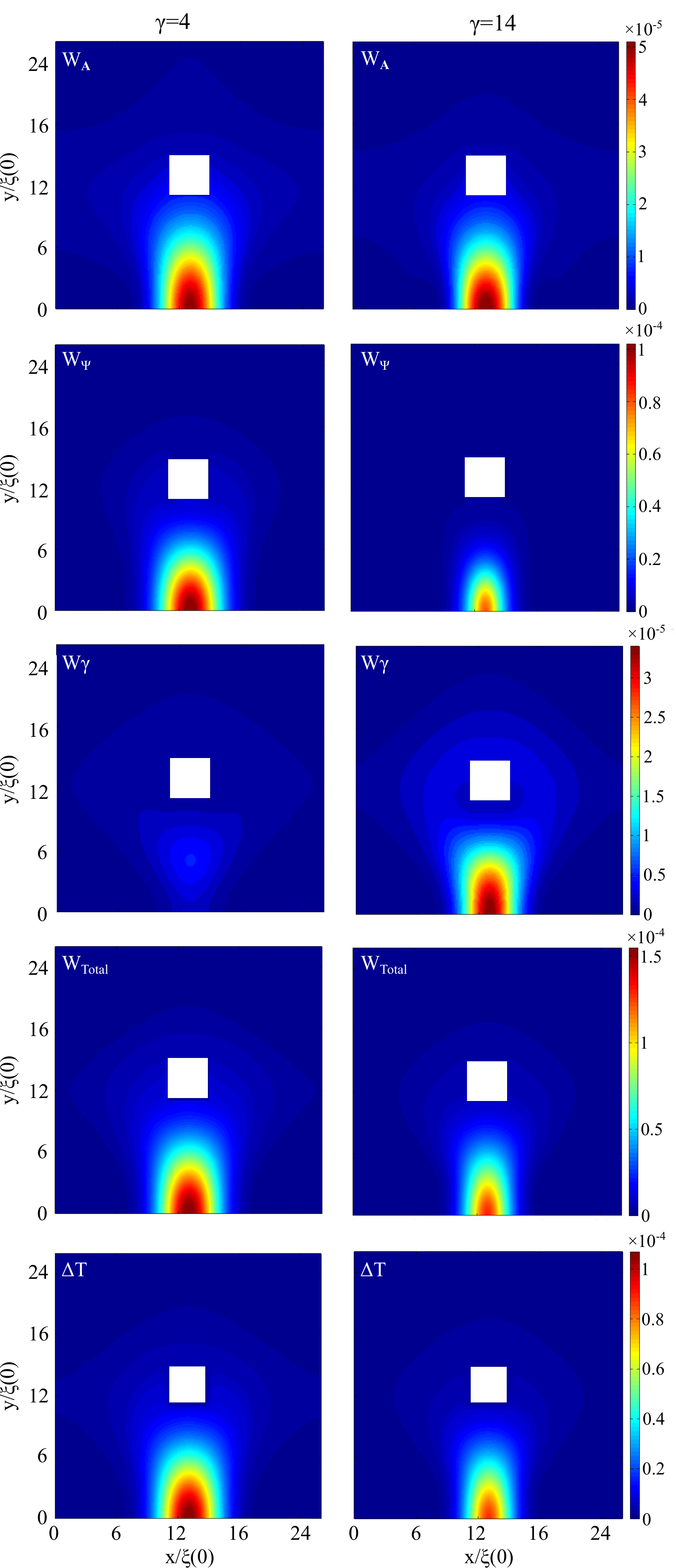}
	\caption{\label{fig:Figure10}Plots of the the local dissipative mechanisms, ${W}_{\mathbf{A}}$, ${W}_{\Psi}$, ${W}_{\gamma}$, ${W}_{total}$ and the local variation of temperature, $\Delta{T}$, for $\gamma=4$ (left column) and $\gamma=14$ (right column). The snapshots were taken from the moment when the vortex is about to leave the sample.}
\end{figure}

In Fig.~\ref{fig:Figure10} we present snapshots of the three dissipative mechanisms and their sum, ${W}_{total}$, and $\Delta{T}$ for $\gamma = 4$ and $\gamma = 14$.
Following the trend discussed about Fig.~\ref{fig:Figure9}, one sees that ${W}_{\Psi}$ is decreased with $\gamma$ while ${W}_{\mathbf{A}}$ increases. In addition, the shading effect over ${W}_{\bold A}$ is also noticeable and less present in the other terms, similar to the gapless system. One can also notice that $\Delta{T}$ also is less pronounced next to the corners and it is clearly greater for $\gamma = 4$ than for $\gamma = 14$.

\begin{figure}[t]
	\centering\vspace*{0.3cm}
	\includegraphics[width=\linewidth]{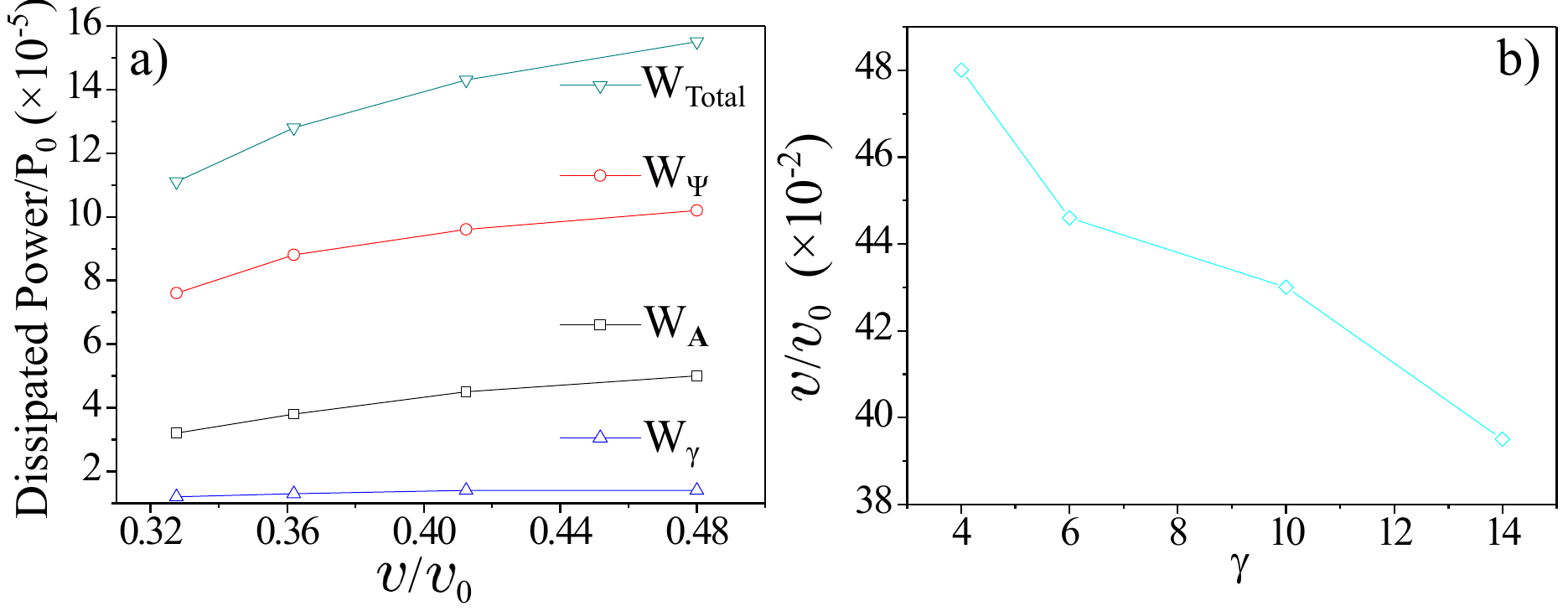}
	\caption{\label{fig:Figure11}(a) The three dissipative mechanisms and their sum, ${W}_{total}$, as a function of the vortex velocity for $\gamma = 4$; (b) vortex velocity (taken just before it left the sample) as function of $\gamma$ for the sample ${L} = 26\xi(0)$ and ${w} = 4\xi(0)$.}
\end{figure}

By analyzing the local maximum dissipated power as a function of the vortex velocity for $\gamma = 4$, shown in Fig.~\ref{fig:Figure11}-(a), it can be seen that ${W}_{\mathbf{A}}$ and ${W}_{\Psi}$ increase with the velocity, while  ${W}_{\gamma}$ stays nearly constant. Besides, the vortex velocity decreases with $\gamma$, as seen in Fig.~\ref{fig:Figure11}(b), which can be explained roughly by the presence of $\gamma$ in the denominator multiplying the time-derivative in Eq.~\ref{eq:1}. In fact, this means an increase in the effective viscosity with $\gamma$.

\begin{figure}[t]
	\centering
	\includegraphics[width=1\columnwidth,height=0.8\linewidth]{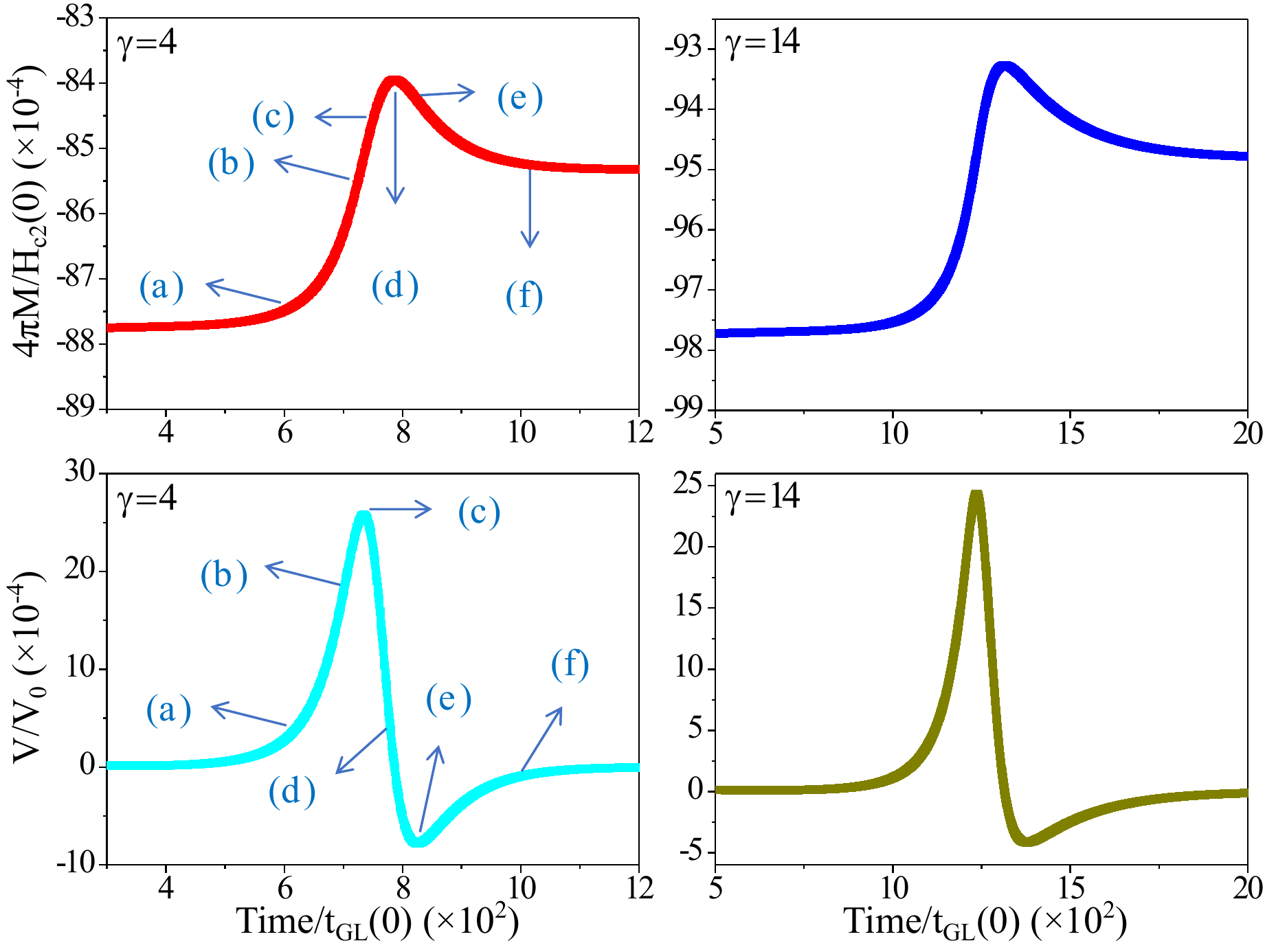}
	\caption{\label{fig:Figure12}Magnetization and voltage as function of the time during the process of one vortex leaving the sample for $\gamma = 4$ (left column) and $\gamma = 14$ (right column).}
\end{figure}

\begin{figure}[t]
	\centering
	\includegraphics[width=\linewidth]{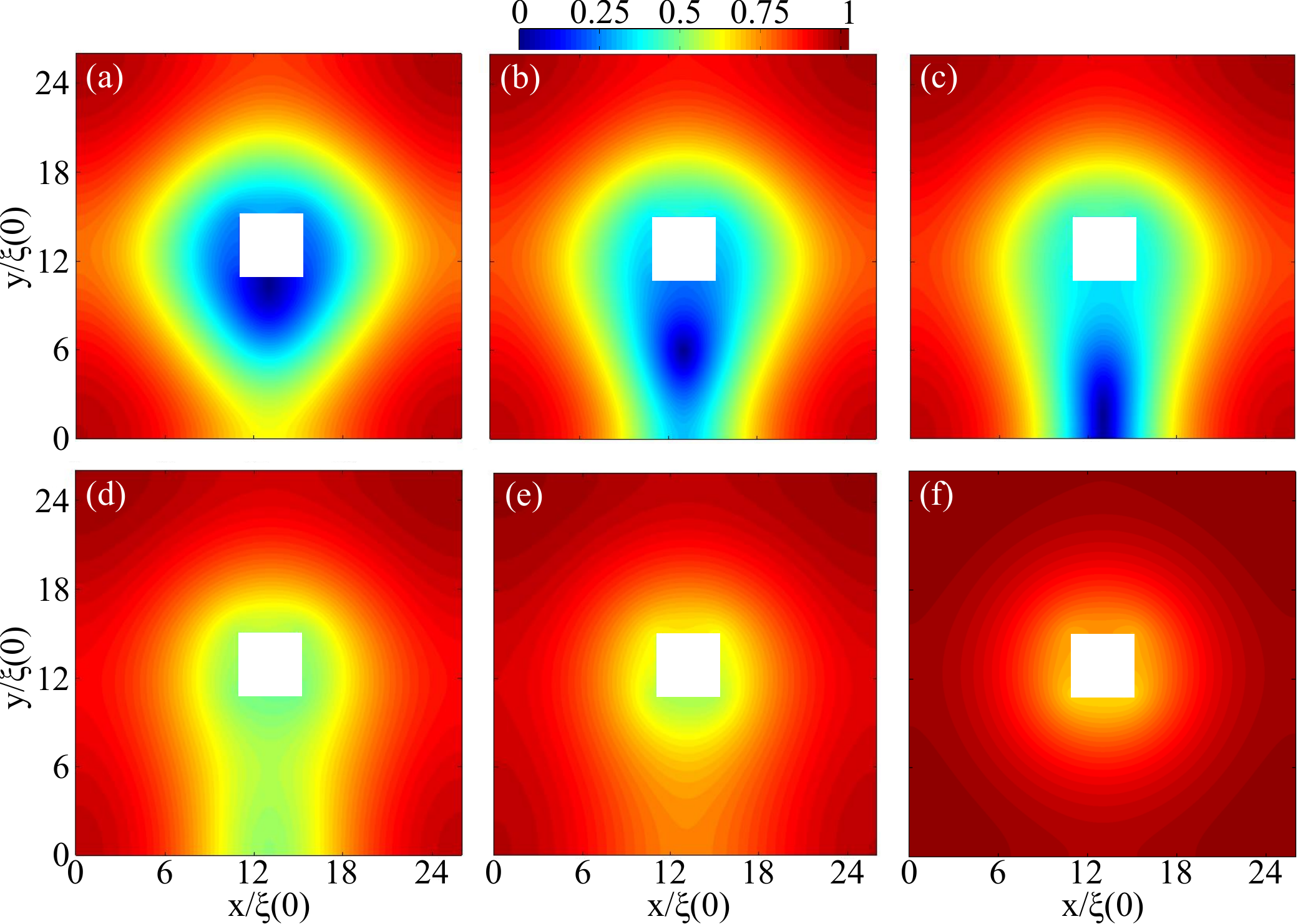}
	\caption{\label{fig:Figure13}Snapshots of $|\Psi|$ exhibiting the vortex dynamics for $\gamma = 4$ at instants (a) to (f) indicated in Figure \ref{fig:Figure12}.}
\end{figure}

Continuing the study of gap systems, we plotted the time-evolution of both magnetization ${M}(t)$ and induced voltage ${V}(t)$ during the vortex motion. The influence of $\gamma$ is qualitatively small, as shown in Fig.~\ref{fig:Figure12}, with only a slight change in the amplitude of the curves. The magnetization of the sample starts from a negative value and ends in a smaller value in modulus after the vortex leaves the sample. This whole process can be easily understood under the scope of the Amp\`{e}re law, where essentially a variation of the magnetic flux over the sample is related to a voltage around it.
The series of 6 points selected along the curves starts with the vortex leaving the occlusion in the center and piercing the film, (a), until the vortex leaving the sample, (f). These snapshots are shown in Fig.~\ref{fig:Figure13}. The local voltage maximum occurs in instants just before a vortex leaves the sample, seen in Figure~\ref{fig:Figure13}-(c). In opposition, the maximum magnetization occurs for the vortex leaving the sample (close to $V(t) = 0$), as it is shown in Fig.~\ref{fig:Figure13}-(d). Also, a negative voltage peak indicates a recovering of the superconducting state or, in other words, an increasing of the modulus of the order parameter in the lower edge of the sample after the vortex leaves the sample, seen in the sequence from~\ref{fig:Figure13}-(d) to (f).

\section{Discussion of the results and conclusions}\label{disconc}
Our study of the resistive state in a gapless superconductor shows that the dissipative mechanisms and the induced voltage are $u$-dependent. Concerning such an issue, Vodolazov and Peeters previously verified that ${u}$ plays an important role, but solely on the relaxation of $\Psi$~\cite{vodolazov2000effect}. Nonetheless, the magnetization does not present a considerable variation by changing $u$. For the clean limit (or with small paramagnetic impurities, i.e., $u=5.79$), the superconducting systems present significant differences between ${W}_{\mathbf{A}}$, ${W}_{\Psi}$, and $V$ for all sizes of the studied samples. Furthermore, it was showed that for $20\xi(0) \le {L}\le 40\xi(0)$, ${W}_{total}$ is nearly the same for both ${u} = 5.79$ and ${u} =12$.

\begin{figure}[t]
	\centering\vspace*{0.3cm}
	\includegraphics[width=1\columnwidth,height=1\linewidth]{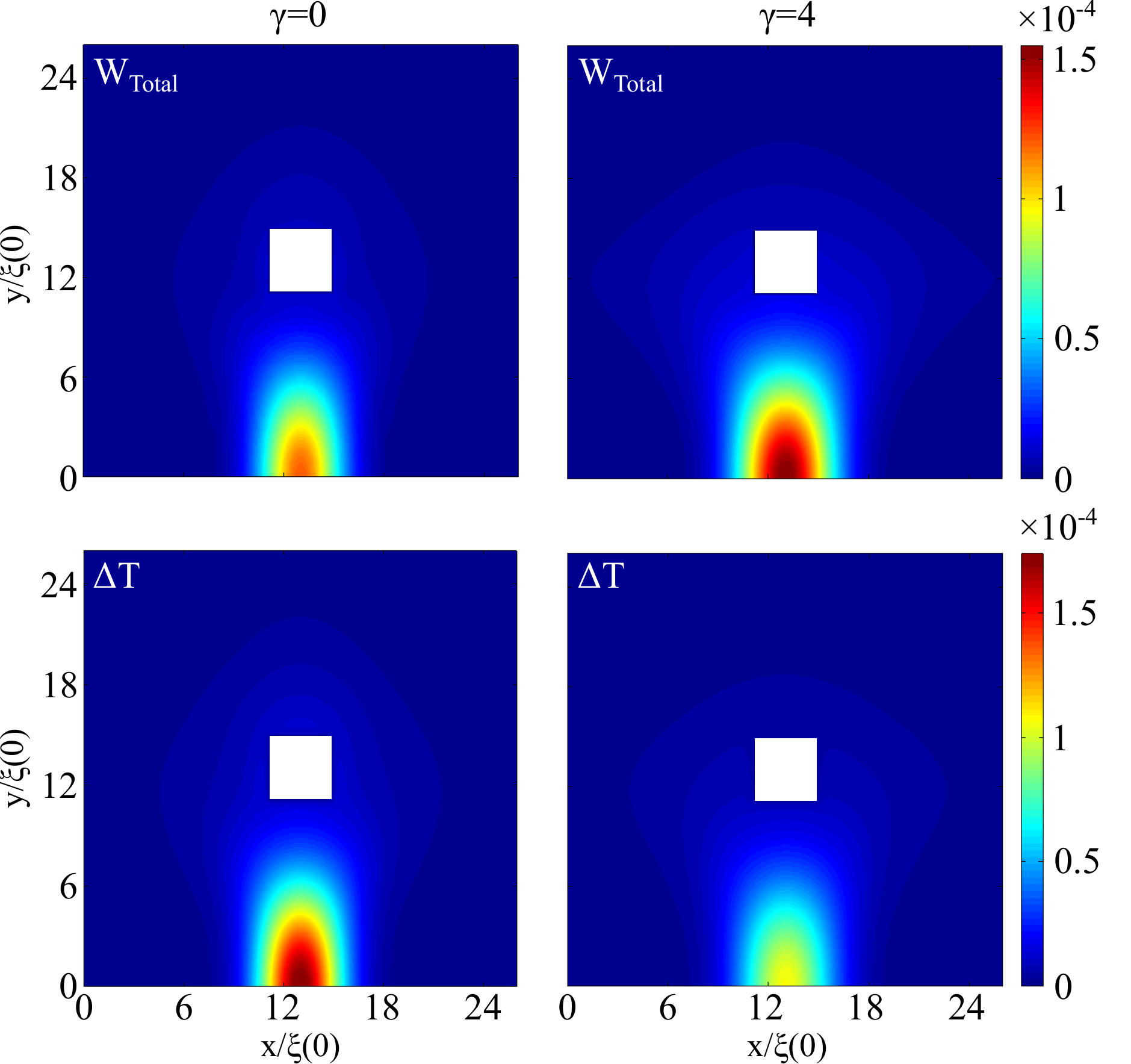}
	\caption{\label{fig:Figure14}Comparing ${W}_{total}$ and $\Delta{T}$ between a gap $(\gamma = 4)$ and a gapless $(\gamma = 0)$ superconductor for ${L} = 26\xi(0)$ and ${w} = 4\xi(0)$.}
\end{figure}

Except ${W}_{\gamma}$, all the other dissipative mechanisms are sensitive to the vortex velocity for both gap and gapless superconductors. ${W}_{\Psi}$ and ${W}_{\gamma}$ presented significant variations with $\gamma$ as well as the shape of the vortex core. In Fig.~\ref{fig:Figure14}, ${W}_{total}$, and $\Delta{T}$ were compared for both gapless and gap systems by considering a sample with ${L} = 26\xi(0)$, and ${w} = 4\xi(0)$ at ${T}_{0} = 0.93{T}_{c}$.
Despite $W_{total}$ in a gap sample being greater than in a gapless one, $\Delta{T}$ is higher for gapless superconductors. Such a fact is due to the ${u}$-dependence of ${C}_{eff}$ and ${K}_{eff}$. Then, in the numerical solution of the TDGL equations coupled with the diffusion equation, $\Delta{T}$ increases proportionally with ${u}$. In addition, it is worth noticing that in our simulations for $\gamma= 4, 6, 10,$ and $14$ we obtained $\Delta{T}_{gap} > \Delta{T}_{gapless}$ and ${W}_{total}^{gap}/{W}_{total}^{gapless}$ $\approx$ ${u}_{gapless}/{{u}_{gap}} = 12/5.79 = 2.07$.

It was shown that ${W}_{\Psi}$ governs the dissipation of Abrikosov vortex dynamics, and it is of fundamental importance for the evaluation of $\Delta{T}$. Besides, other theoretical works showed that ${W}_{total}$ depends on the $\Psi$ relaxation for ${H} << {H}_{c2}$ \cite{gorkov1971viscous,kupriyanov1972}. Kupriyanov and Likharev demonstrated that $\frac{\Gamma_{R}}{\Gamma_{N}} =  1.7547$, is the rate between the normal electron relaxation time ($\Gamma_{N}$) and the $\Psi$ relaxation time for gapless superconductors ($\Gamma_{R}$). 
Considering $\Gamma_{N}$ related to $W_{\Psi}$ and $\Gamma_{R}$ related to ${W}_{\mathbf{A}}$, one obtained $1.61 <\frac{{W}_{\Psi}}{{W}_{\mathbf{A}}} < 2.19$ (from Figure 5), which is in accordance with~\cite{kupriyanov1972}. On the other hand, for the gap 
systems ($\gamma = 4$) the ratio $2.04 < \frac{{W}_{\Psi}}{{W}_{\mathbf{A}}} < 2.37$ (from Figure 8) and therefore, our results agree with Ref.~\cite{kupriyanov1972}, i.e. ${W}_{\mathbf{A}}$ is essentially not affected by $\gamma$ while ${W}_{\Psi}$ increases. As found in the literature, resistivity is related to the dissipated power during vortex motion. However, just a few works compare experimental results with the theoretical ones at ${H} << {H}_{c2}$. In the works of Fogel~\cite{fogel1973temperature}, and Gubankov \textit{et al.}~\cite{gubankov1973resistance}, it was shown that the viscosity is lower than that one verified experimentally by considering dissipation just caused by normal electrons.

However, by taking into account the power dissipation due to relaxation of $\Psi$, an excellent experimental agreement is reached \cite{fogel1973temperature}. For a thin film with a small concentration of paramagnetic impurities, the process involving the relaxation of $\Psi$ is also the main dissipative mechanism \cite{takayama1977flux}. Poon and Wong showed that, by measuring the electrical properties of Zr$_3$Ni and Zr$_3$Rh bulk samples, it is not possible to obtain an experimental agreement by considering just one type of dissipative mechanism. Nonetheless, when all dissipative mechanisms are considered, it is possible to obtain an excellent agreement with experimental data. Therefore, the contributions of the relaxation terms and the normal currents are of great importance and agree well with experiments in the temperature range $0.66 {T}_{c} < {T}_{0} < 0.95{T}_{c}$ \cite{poon1983viscous}. It is important to emphasize that the experimental data in previous studies ~\cite{fogel1973temperature,takayama1977flux,poon1983viscous}, agreed well with  those ones from theoretical works \cite{bardeen1965theory,tinkham1964viscous,gorkov1971viscous,kupriyanov1972}, where the dissipative mechanisms were obtained in a different formalism than we presented here. In this work, all dissipative mechanisms were derived from the generalization of the free energy theorem \cite{schmid1966time,duarte2017dynamics}. Thus, all dissipative mechanisms that are experimentally relevant for an accurate description of metallic superconducting materials were considered. It is worth mentioning that the values obtained for $\Delta{T}$ are of the order of $10^{-4} {T}_{c}$ and are suitable to be measured with nanoSQUIDs as it was demonstrated by Halbertal \textit{et al.}\ \cite{halbertal2016nanoscale}. For the ${H} << {H}_{c2}$ regime, one demonstrated that ${W}_{\Psi}$ is of great importance to ${W}_{total}$, and consequently to $\Delta{T}$. Furthermore, the vortex velocity and the nanostructured superconductors geometry  influence the values of ${W}_{\Psi}$. For gap superconductors, it was shown that lower $\gamma$ ($\leq 4$) increases ${W}_{\Psi}$.

In conclusion, our theoretical approach aims to describe dissipative and thermal diffusion processes in superconducting devices such as single-photon detectors, fluxonic cellular automata \cite{milovsevic2007fluxonic,milovsevic2010vortex}, or vortex random access memory \cite{golod2015single}, where the control of the dissipation during vortex motion is very important to reach a reliable performance. The characteristic times obtained for vortex motion are in accordance with experimental results for the characteristic times 

We acknowledge the support from the Brazilian agencies: S\~{a}o Paulo Research Foundation, FAPESP, grants 2016/12390-6 and 2020/10058-0, Coordena\c c\~ao de Aperfei\c coamento de Pessoal de N\'ivel Superior - Brasil (CAPES) - Finance Code 001, and National Council of Scientific and Technological Development (CNPq, grant 310428/2021-1) for financial support. The work at HSE University (T.T.S. and A.S.V.)
was supported by the framework of the Academic Fund Program at HSE University in 2021 (Grant No. 21-04-041).

\bibliographystyle{unsrt}
\bibliography{ref}

\begin{thebibliography}{10}

\bibitem{gol2001picosecond}
G~N Gol’Tsman, O~Okunev, G~Chulkova, A~Lipatov, A~Semenov, K~Smirnov,
  B~Voronov, A~Dzardanov, C~Williams, and R~Sobolewski.
\newblock Picosecond superconducting single-photon optical detector.
\newblock {\em Applied physics letters}, 79(6):705--707, 2001.

\bibitem{kerman2007constriction}
A~J Kerman, E~A Dauler, J~K~W Yang, Kristine~M Rosfjord, V~Anant, K~K Berggren,
  G~N Gol’tsman, and B~M Voronov.
\newblock Constriction-limited detection efficiency of superconducting nanowire
  single-photon detectors.
\newblock {\em Applied Physics Letters}, 90(10):101110, 2007.

\bibitem{dorenbos2008low}
S~N Dorenbos, E~M Reiger, U~Perinetti, V~Zwiller, T~Zijlstra, and T~M Klapwijk.
\newblock Low noise superconducting single photon detectors on silicon.
\newblock {\em Applied Physics Letters}, 93(13):131101, 2008.

\bibitem{Hadfield2009}
Robert~H. Hadfield.
\newblock Single-photon detectors for optical quantum information applications.
\newblock {\em Nature Photonics}, 3(12):696--705, Dec 2009.

\bibitem{berdiyorov2012spatially}
G~R Berdiyorov, M~V Milo{\v{s}}evi{\'c}, and F~M Peeters.
\newblock Spatially dependent sensitivity of superconducting meanders as
  single-photon detectors.
\newblock {\em Applied Physics Letters}, 100(26):262603, 2012.

\bibitem{zotova2012photon}
A~N Zotova and D~Yu Vodolazov.
\newblock Photon detection by current-carrying superconducting film: A
  time-dependent ginzburg-landau approach.
\newblock {\em Physical Review B}, 85(2):024509, 2012.

\bibitem{Natarajan2012}
Chandra~M Natarajan, Michael~G Tanner, and Robert~H Hadfield.
\newblock Superconducting nanowire single-photon detectors: physics and
  applications.
\newblock {\em Superconductor Science and Technology}, 25(6):063001, apr 2012.

\bibitem{gaudio2014inhomogeneous}
R~Gaudio, K~P~M Op't~Hoog, Z~Zhou, D~Sahin, and A~Fiore.
\newblock Inhomogeneous critical current in nanowire superconducting
  single-photon detectors.
\newblock {\em Applied Physics Letters}, 105(22):222602, 2014.

\bibitem{renema2015effect}
J~J Renema, R~J Rengelink, I~Komen, Q~Wang, R~Gaudio, K~P~M op't Hoog, Z~Zhou,
  D~Sahin, A~Fiore, P~Kes, et~al.
\newblock The effect of magnetic field on the intrinsic detection efficiency of
  superconducting single-photon detectors.
\newblock {\em Applied Physics Letters}, 106(9):092602, 2015.

\bibitem{rosticher2010high}
M~Rosticher, F~R Ladan, J~P Maneval, S~N Dorenbos, T~Zijlstra, T~M Klapwijk,
  V~Zwiller, A~Lupa{\c{s}}cu, and G~Nogues.
\newblock A high efficiency superconducting nanowire single electron detector.
\newblock {\em Applied Physics Letters}, 97(18):183106, 2010.

\bibitem{Semenov2002}
Alexei~D Semenov, Gregory N~Gol tsman, and Roman Sobolewski.
\newblock 15(4):R1--R16, mar 2002.

\bibitem{Kozerov2000}
A.~G. Kozorezov, A.~F. Volkov, J.~K. Wigmore, A.~Peacock, A.~Poelaert, and
  R.~den Hartog.
\newblock Quasiparticle-phonon downconversion in nonequilibrium
  superconductors.
\newblock {\em Phys. Rev. B}, 61:11807--11819, May 2000.

\bibitem{yang2007modeling}
J~K~W Yang, A~J Kerman, E~A Dauler, V~Anant, K~M Rosfjord, and K~K Berggren.
\newblock Modeling the electrical and thermal response of superconducting
  nanowire single-photon detectors.
\newblock {\em IEEE transactions on applied superconductivity}, 17(2):581--585,
  2007.

\bibitem{Abrikosov1957a}
A.~A. Abrikosov.
\newblock On the magnetic properties of superconductors of the second group.
\newblock {\em J. Exptl. Theoret. Phys.}, 32:1174--1182, 1957.

\bibitem{Brandt1995}
E~H Brandt.
\newblock The flux-line lattice in superconductors.
\newblock {\em Reports on Progress in Physics}, 58(11):1465--1594, nov 1995.

\bibitem{TinkhamBook}
Michael Tinkham.
\newblock {\em Introduction to superconductivity}.
\newblock McGraw-Hill New York, 1975.

\bibitem{berdiyorov2012magnetoresistance}
G~R Berdiyorov, X~H Chao, F~M Peeters, H~B Wang, V~V Moshchalkov, and B~Y Zhu.
\newblock Magnetoresistance oscillations in superconducting strips: a
  ginzburg-landau study.
\newblock {\em Physical Review B}, 86(22):224504, 2012.

\bibitem{berdiyorov2012large}
G~R Berdiyorov, M~V Milo{\v{s}}evi{\'c}, M~L Latimer, Z~L Xiao, W~K Kwok, and
  F~M Peeters.
\newblock Large magnetoresistance oscillations in mesoscopic superconductors
  due to current-excited moving vortices.
\newblock {\em Physical review letters}, 109(5):057004, 2012.

\bibitem{hernandez2008dissipation}
A~D Hern{\'a}ndez and D~Dom{\'\i}nguez.
\newblock Dissipation spots generated by vortex nucleation points in mesoscopic
  superconductors driven by microwave magnetic fields.
\newblock {\em Physical Review B}, 77(22):224505, 2008.

\bibitem{bardeen1965theory}
J~Bardeen and M~J Stephen.
\newblock Theory of the motion of vortices in superconductors.
\newblock {\em Physical Review}, 140(4A):A1197, 1965.

\bibitem{tinkham1964viscous}
M~Tinkham.
\newblock Viscous flow of flux in type-{II} superconductors.
\newblock {\em Physical Review Letters}, 13(26):804, 1964.

\bibitem{Gerhenzon1984}
E.M. Gershenzon, M.E. Gershenzon, G.N. Gol'tsman, A.D. Semyonov, and A.V.
  Sergeev.
\newblock Heating of electrons in superconductor in the resistive state due to
  electromagnetic radiation.
\newblock {\em Solid State Communications}, 50(3):207--212, 1984.

\bibitem{Gershenzon1990}
E.M. Gershenzon, G.N. Gol'tsman, A.D. Semenov, and A.V. Sergeev.
\newblock Mechanism of picosecond response of granular ybacuo films to
  electromagnetic radiation.
\newblock {\em Solid State Communications}, 76(4):493--497, 1990.

\bibitem{Marsili2016}
F.~Marsili, M.~J. Stevens, A.~Kozorezov, V.~B. Verma, Colin Lambert, J.~A.
  Stern, R.~D. Horansky, S.~Dyer, S.~Duff, D.~P. Pappas, A.~E. Lita, M.~D.
  Shaw, R.~P. Mirin, and S.~W. Nam.
\newblock Hotspot relaxation dynamics in a current-carrying superconductor.
\newblock {\em Phys. Rev. B}, 93:094518, Mar 2016.

\bibitem{Sheikhzada2020}
Ahmad Sheikhzada and Alex Gurevich.
\newblock Dynamic pair-breaking current, critical superfluid velocity, and
  nonlinear electromagnetic response of nonequilibrium superconductors.
\newblock {\em Phys. Rev. B}, 102:104507, Sep 2020.

\bibitem{Tanaka1996}
Ken’ichi Tanaka, Yoshiro Arikawa, Matsuo Sekine, Motoichi Ohtsu, Yuichi
  Harada, and Martin Danerud.
\newblock Highly sensitive and wideband optical detection in patterned
  {YB}a$_2${C}u$_3${O}$_{7-\delta}$ thin films.
\newblock {\em Applied Physics Letters}, 68(22):3174--3176, 1996.

\bibitem{Bitauld2010}
David Bitauld, Francesco Marsili, Alessandro Gaggero, Francesco Mattioli,
  Roberto Leoni, Saeedeh~Jahanmiri Nejad, Francis L{\'e}vy, and Andrea Fiore.
\newblock Nanoscale optical detector with single-photon and multiphoton
  sensitivity.
\newblock {\em Nano Letters}, 10(8):2977--2981, Aug 2010.

\bibitem{schmid1966time}
A~Schmid.
\newblock A time dependent ginzburg-landau equation and its application to the
  problem of resistivity in the mixed state.
\newblock {\em Physik der kondensierten Materie}, 5(4):302--317, 1966.

\bibitem{kramer1978theory}
L~Kramer and R~J Watts-Tobin.
\newblock Theory of dissipative current-carrying states in superconducting
  filaments.
\newblock {\em Physical Review Letters}, 40(15):1041, 1978.

\bibitem{duarte2017dynamics}
ECS Duarte, E~Sardella, WA~Ortiz, and R~Zadorosny.
\newblock Dynamics and heat diffusion of abrikosov’s vortex-antivortex pairs
  during an annihilation process.
\newblock {\em Journal of Physics: Condensed Matter}, 29(40):405605, 2017.

\bibitem{petkovic2016deterministic}
I~Petkovi{\'c}, A~Lollo, L~I Glazman, and J~G~E Harris.
\newblock Deterministic phase slips in mesoscopic superconducting rings.
\newblock {\em Nature communications}, 7(1):1--7, 2016.

\bibitem{vodolazov2005masking}
D~Yu Vodolazov, F~M Peeters, M~Morelle, and V~V Moshchalkov.
\newblock Masking effect of heat dissipation on the current-voltage
  characteristics of a mesoscopic superconducting sample with leads.
\newblock {\em Physical Review B}, 71(18):184502, 2005.

\bibitem{silhanek2010formation}
A~V Silhanek, M~V Milo{\v{s}}evi{\'c}, R~B~G Kramer, G~R Berdiyorov, J~Van~de
  Vondel, R~F Luccas, T~Puig, F~M Peeters, and V~V Moshchalkov.
\newblock Formation of stripelike flux patterns obtained by freezing kinematic
  vortices in a superconducting pb film.
\newblock {\em Physical review letters}, 104(1):017001, 2010.

\bibitem{jelic2015stroboscopic}
{\v{Z}}~L Jeli{\'c}, M~V Milo{\v{s}}evi{\'c}, J~Van~de Vondel, and A~V
  Silhanek.
\newblock Stroboscopic phenomena in superconductors with dynamic pinning
  landscape.
\newblock {\em Scientific reports}, 5:14604, 2015.

\bibitem{jelic2016velocimetry}
{\v{Z}}~L Jeli{\'c}, M~V Milo{\v{s}}evi{\'c}, and A~V Silhanek.
\newblock Velocimetry of superconducting vortices based on stroboscopic
  resonances.
\newblock {\em Scientific reports}, 6:35687, 2016.

\bibitem{gor1975vortex}
L~P Gor'kov and N~B Kopnin.
\newblock Vortex motion and resistivity of type-ll superconductors in a
  magnetic field.
\newblock {\em Soviet Physics Uspekhi}, 18(7):496, 1975.

\bibitem{kramer1977lossless}
L~Kramer and A~Baratoff.
\newblock Lossless and dissipative current-carrying states in
  quasi-one-dimensional superconductors.
\newblock {\em Physical Review Letters}, 38(9):518, 1977.

\bibitem{gropp1996numerical}
W~D Gropp, H~G Kaper, G~K Leaf, D~M Levine, M~Palumbo, and V~M Vinokur.
\newblock Numerical simulation of vortex dynamics in type-{II} superconductors.
\newblock {\em Journal of Computational Physics}, 123(2):254--266, 1996.

\bibitem{milovsevic2010ginzburg}
M~V Milo{\v{s}}evi{\'c} and R~Geurts.
\newblock The ginzburg--landau theory in application.
\newblock {\em Physica C: Superconductivity}, 470(19):791--795, 2010.

\bibitem{sardella2006temperature}
E~Sardella, A~L Malvezzi, P~N Lisboa-Filho, and W~A Ortiz.
\newblock Temperature-dependent vortex motion in a square mesoscopic
  superconducting cylinder: Ginzburg-landau calculations.
\newblock {\em Physical Review B}, 74(1):014512, 2006.

\bibitem{vodolazov2000effect}
D~Yu Vodolazov.
\newblock Effect of surface defects on the first field for vortex entry in
  type-{II} superconductors.
\newblock {\em Physical Review B}, 62(13):8691, 2000.

\bibitem{tidecks1986continuous}
R~Tidecks and Th~Werner.
\newblock Continuous change from weak to strong coupling behavior in
  quasi-one-dimensional superconductors.
\newblock {\em Journal of low temperature physics}, 65(3-4):151--184, 1986.

\bibitem{gorkov1971viscous}
L~P Gorkov and N~B Kopnin.
\newblock Viscous vortex flow in superconductors with paramagnetic impurities.
\newblock {\em J. Exp. Theor. Phys.}, 33(6):1251--1256, 1971.

\bibitem{baranov2011current}
V~V Baranov, A~G Balanov, and V~V Kabanov.
\newblock Current-voltage characteristic of narrow superconducting wires:
  Bifurcation phenomena.
\newblock {\em Physical Review B}, 84(9):094527, 2011.

\bibitem{ivlev1980dynamics}
B~I Ivlev, N~B Kopnin, and L~A Maslova.
\newblock Dynamics of the resistive state of a superconductor.
\newblock {\em Sov. Phys.-JETP (Engl. Transl.);(United States)}, 51(5), 1980.

\bibitem{ivlev1985low}
B~I Ivlev, N~B Kopnin, and I~A Larkin.
\newblock Low-frequency oscillations in the resistive state of narrow
  superconductors.
\newblock {\em Sov. Phys. JETP}, 61(2):337--343, 1985.

\bibitem{zadorosny2012crossover}
R~Zadorosny, E~Sardella, A~L Malvezzi, P~N Lisboa-Filho, and W~A Ortiz.
\newblock Crossover between macroscopic and mesoscopic regimes of vortex
  interactions in type-{II} superconductors.
\newblock {\em Physical Review B}, 85(21):214511, 2012.

\bibitem{Zhang2018}
Xiaofu Zhang, Adriana~E. Lita, Mariia Sidorova, Varun~B. Verma, Qiang Wang,
  Sae~Woo Nam, Alexei Semenov, and Andreas Schilling.
\newblock Superconducting fluctuations and characteristic time scales in
  amorphous {WSi}.
\newblock {\em Phys. Rev. B}, 97:174502, May 2018.

\bibitem{berdiyorov2009kinematic}
G.~R. {Berdiyorov}, M.~V. {Milošević}, and F.~M. {Peeters}.
\newblock Kinematic vortex-antivortex lines in strongly driven superconducting
  stripes.
\newblock {\em Physical Review B}, 79(18), 2009.

\bibitem{Brosens1999}
F.~Brosens, J.T. Devreese, V.M. Fomin, and V.V. Moshchalkov.
\newblock Superconductivity in a wedge: analytical variational results.
\newblock {\em Solid State Communications}, 111(10):565--569, 1999.

\bibitem{Sidorova2018}
Mariia~V. Sidorova, A.~G. Kozorezov, A.~V. Semenov, Yu.~P. Korneeva, M.~Yu.
  Mikhailov, A.~Yu. Devizenko, A.~A. Korneev, G.~M. Chulkova, and G.~N.
  Goltsman.
\newblock Nonbolometric bottleneck in electron-phonon relaxation in ultrathin
  wsi films.
\newblock {\em Phys. Rev. B}, 97:184512, May 2018.

\bibitem{Zhang2019}
Lu~Zhang, Lixing You, Xiaoyan Yang, Yan Tang, Mengting Si, Kaixin Yan, Weijun
  Zhang, Hao Li, Hui Zhou, Wei Peng, and Zhen Wang.
\newblock Hotspot relaxation time in disordered niobium nitride films.
\newblock {\em Applied Physics Letters}, 115(13):132602, 2019.

\bibitem{kupriyanov1972}
M~Yu Kupriyanov and K~K Likharev.
\newblock {\em ZhETF Pis. Red.}, 15, 1972.

\bibitem{fogel1973temperature}
N~Fogel.
\newblock Temperature dependence of the viscosity coefficient in type {II}
  superconductors.
\newblock {\em Sov. Phys. JETP}, 36(4):725--730, 1973.

\bibitem{gubankov1973resistance}
V~N Gubankov.
\newblock Resistance of thin superconducting films in the dynamic mixed state.
\newblock Technical report, Institute of Radio Technology and Electronics,
  Academy of Sciences of the Soviet Union, 1973.

\bibitem{takayama1977flux}
T~Takayama.
\newblock Flux flow resistivity of superconducting thin film.
\newblock {\em Journal of Low Temperature Physics}, 27(3-4):359--396, 1977.

\bibitem{poon1983viscous}
S~J Poon and K~M Wong.
\newblock Viscous flow of vortices in ideal type-{II} amorphous
  superconductors.
\newblock {\em Physical Review B}, 27(11):6985, 1983.

\bibitem{halbertal2016nanoscale}
D~Halbertal, J~Cuppens, M~B Shalom, L~Embon, N~Shadmi, Y~Anahory, H~R Naren,
  J~Sarkar, A~Uri, Y~Ronen, et~al.
\newblock Nanoscale thermal imaging of dissipation in quantum systems.
\newblock {\em Nature}, 539(7629):407--410, 2016.

\bibitem{milovsevic2007fluxonic}
M~V Milo{\v{s}}evi{\'c}, G~R Berdiyorov, and F~M Peeters.
\newblock Fluxonic cellular automata.
\newblock {\em Applied Physics Letters}, 91(21):212501, 2007.

\bibitem{milovsevic2010vortex}
M~V Milo{\v{s}}evi{\'c} and F~M Peeters.
\newblock Vortex manipulation in a superconducting matrix with view on
  applications.
\newblock {\em Applied Physics Letters}, 96(19):192501, 2010.

\bibitem{golod2015single}
T~Golod, A~Iovan, and V~M Krasnov.
\newblock Single abrikosov vortices as quantized information bits.
\newblock {\em Nature communications}, 6(1):1--5, 2015.

\end{thebibliography}
\end{document}